\documentclass[iop, revtex4]{emulateapj}

\newcommand{\Msol}{\mbox{$M_\odot$}} 
 
\newcommand{\kms}{\mbox{km s$^{-1}$}}

\newcommand{\hr}{\mbox{$^{\rm h}$}} 
\newcommand{\mn}{\mbox{$^{\rm m}$}}

\usepackage[usenames]{color}
\usepackage{multirow}
\usepackage{lscape} 
\slugcomment{Not to appear in Nonlearned J., 45.}

\shorttitle{CO(2--1) in NGC 5253 }
\shortauthors{Rie E. Miura et al.}

\begin{document}


\title{ALMA observations toward the Starburst Dwarf Galaxy NGC\,5253: {\sc i}. Molecular Cloud properties and scaling relations}


\author{Rie E. Miura\altaffilmark{1}}
%\affil{NAOJ}
\email{rie.miura@nao.ac.jp}

\author{Daniel Espada\altaffilmark{1,2}}

\author{Akihiko Hirota\altaffilmark{1,3}}

\author{Kouichiro Nakanishi\altaffilmark{1,2}}

\author{George J. Bendo\altaffilmark{4}}

\author{Hajime Sugai\altaffilmark{5}}

\altaffiltext{1}{National Astronomical Observatory of Japan, National Institutes of Natural Sciences, 2-21-1 Osawa, Mitaka, Tokyo 181-8588, Japan}
\altaffiltext{2}{Department of Astronomical Science, The Graduate University for Advanced Studies (Sokendai), 2-21-1 Osawa, Mitaka, Tokyo 181-0015, Japan}
\altaffiltext{3}{Joint ALMA Observatory, Alonso de Cordova 3107, Vitacura 763-0355, Santiago de Chile}
\altaffiltext{4}{UK ALMA Regional Centre Node, Jodrell Bank Centre for Astrophysics, School of Physics and Astronomy, The University of Manchester, Oxford Road, Manchester M13 9PL, United Kingdom}
\altaffiltext{5}{Kavli Institute for the Physics and Mathematics of the Universe, Todai Institutes for Advanced Study, The University of Tokyo (Kavli IPMU, WPI), 5-1-5 Kashiwanoha, Kashiwa, Chiba 277-8583, Japan}

\begin{abstract}
We present high-spatial-resolution ($\sim 0\farcs2$, or $\sim$3\,pc) CO(2--1) observations of the nearest young starburst dwarf galaxy, NGC\,5253, taken with the Atacama Large Millimeter/submillimeter Array. 
We have identified 118 molecular clouds with average values of 4.3\,pc in radius and 2.2\,\kms\, in velocity dispersion, which comprise the molecular cloud complexes observed previously with $\sim$100\,pc resolution.
We derive for the first time in this galaxy the $I{\rm (CO)}$--$N$(H$_2$) conversion factor, $X$ = $4.1^{+5.9}_{-2.4}\times10^{20}$\,cm$^{-2}$(K\,\kms)$^{-1}$,  based on the virial method.
The line-width and mass-to-size relations of the resolved molecular clouds present an offset on average toward higher line-widths and masses with respect to quiescent regions in other nearby spiral galaxies and our Galaxy. The offset in the scaling relation reaches its maximum in regions close to the central starburst, where velocity dispersions are $\sim$ 0.5 dex higher and gas mass surface densities are as high as $\Sigma_{\rm H_2}$ = 10$^3$\,\Msol\,pc $^{-2}$.
These central clouds are gravitationally bound despite the high internal pressure. 
A spatial comparison with star clusters found in the literature enables us to identify six clouds that are associated with young star clusters.
Furthermore, the star formation efficiencies (SFEs) of some of these clouds exceed those found in star-cluster-forming clouds within our Galaxy.
We conclude that once a super star cluster is formed, the parent molecular clouds are rapidly dispersed by the destructive stellar feedback, which results in such a high SFE in the central starburst of NGC\,5253.
\end{abstract}

\section{Introduction}
Our knowledge of the properties of molecular clouds is  steadily increasing as more wide-field and high-resolution observations are made available.
Large number statistics of the properties of resolved molecular clouds identified using a similar method now exist for different galaxies  \citep[e.g.][]{2001ApJ...551..852H,2009ApJ...699.1092H,2011ApJS..197...16W,2008ApJ...686..948B,2013ApJ...772..107D,2015ApJ...801...25L,2014ApJ...784....3C,2012ApJ...761...37M,2008ApJS..178...56F}. These studies showed that while the cloud properties and scaling relations are compatible within the Galaxy and external galaxies across a wide range of environments \citep{2008ApJ...686..948B}, those in starburst (SB) regions or in the Galactic Center substantially differ \citep{2001ApJ...562..348O,2015ApJ...801...25L}. For instance, \citet{2015ApJ...801...25L} showed that the clouds in the prototypical SB galaxy NGC\,253 have large line-widths, high surface gas densities, and short free-fall times, which suggest that its efficient star formation is enabled by the surrounding dense gas reservoir.  

An important environment where the different conditions of the molecular gas around SBs can be proved is blue compact dwarf galaxies (BCDs).
BCDs are defined as faint and compact galaxies, and most of them have compact SB cores with high star formation rates \citep{2004AJ....128.2170H}.
These compact SBs comprise a relatively small number of star clusters compared to larger SB galaxies such as NGC\,253, and thus the effect of the SB on the surroundings is less complex. 
Studying gas properties in smaller-scale SBs in BCDs allows us to better understand larger SB systems.

Other dwarf galaxies have in general low gas densities and low star formation efficiencies \citep[SFEs;][]{2010AJ....140.1194B}, which may imply that some mechanism exists in BCDs to change the cloud properties of a typical dwarf galaxy to trigger an SB such as in NGC\,253 (although at smaller scale).
Several external triggering mechanisms to explain the bursts in BCDs have been proposed and explored through numerical simulations, including merger or interaction between gas-rich dwarf galaxies \citep{2008MNRAS.388L..10B} and external gas infall \citep{2014MNRAS.442.1830V}.
These external triggering events can increase the turbulence of the gas and transfer gas from the outer regions to the center, which is prone to be dense enough for gravitational collapse, and then an SB event follows.
These SBs may blow out the gas from the central region, although the outflow gas may fall back and become the seeds of the next SB \citep{2014MNRAS.442.1830V}.

Recent Atacama Large Millimeter/submillimeter Array (ALMA) observations toward the BCD galaxy II Zw 40 BCD have shown that the molecular gas is clumpy and is characterized by larger line-widths and more compact sizes as well as higher molecular gas surface densities with respect to other quiescent molecular clouds in other objects \citep{2016ApJ...828...50K}.
However, still only a small number of observations toward BCDs exist to study the cloud properties under these peculiar environments.

In this paper, we study the nearby\footnote{There are still uncertainties in the distance measurement of $2.8$--$5.2$\,Mpc \citep[e.g.][]{1994ApJ...421L..87B, 2004ApJ...608...42S,1996ApJ...473...88R}. We adopt a distance to NGC\,5253 of 3.15\,Mpc \citep{2001ApJ...553...47F,2007AJ....134.1799D} throughout this paper.} SB galaxy NGC\,5253, which is argued to be a BCD that has experienced \ion{H}{1} gas infall that has triggered its central SB \citep{2012MNRAS.419.1051L}.
This galaxy is known to have a relatively low metallicity \citep[$12+\log({\rm O/H})\sim 8.18$--8.30;][]{2007ApJ...656..168L,1997ApJ...477..679K,1999ApJ...514..544K}.
NGC\,5253 hosts a young nuclear SB, and because of its proximity it has been the object of many studies.
This galaxy has a disk-like morphology extending SE-NW as seen in optical images (see Fig.\,\ref{almafov}, and also Table\,\ref{opt} for a summary of the main galaxy properties) and 
has several young clusters with ages ranging from 1 to 15\,Myr are distributed within the central 300\,pc \citep{2015ApJ...811...75C}.
Older clusters can also be found over the galaxy and their ages are typically 1\,Gyr or more \citep{2013MNRAS.431.2917D}.
The H$\alpha$ image shows multiple filamentary and bubble-like structures extending perpendicular to its optical major axis \citep{2004AJ....127.1405C}.
The multiage nature of the clusters and the existence of bubbles indicate that NGC\,5253 has experienced a few bursts in the past \citep{1998MNRAS.295...43D,2010ApJ...721..297M}.

One of the peculiarities of this galaxy is that its central SB is found to be powered by two very young (1\,Myr) and massive ($\sim3\times10^6\Msol$) compact stellar clusters \citep[so-called super star clusters; hereafter SSC;][]{2015ApJ...811...75C}, separated by 6\,pc \citep{2004ApJ...612..222A,2004A&A...415..509V}.
The SSCs are deeply embedded radio compact (1--2 pc size) \ion{H}{2} regions illuminated by an equivalent of 4000 O stars \citep{2000ApJ...532L.109T}. 
The submillimeter hydrogen recombination line H30$\alpha$ is detected toward the location of the obscured SSCs, and \citet{2017MNRAS.472.1239B} suggest that this region is not only the predominant site of star formation but also potentially accounts for about $\sim$90\% of the total in the nuclear region. 
As a consequence, an unusually high SFE, 37\,\%--75\,\% on a 100 pc scale, is estimated for the central SB region  \citep{2015Natur.519..331T,2015PASJ...67L...1M,2002AJ....124..877M}.

Another peculiarity of the galaxy is the extended atomic gas (\ion{H}{1}), which does not follow galactic rotation even though the galaxy seems to have a stellar disk \citep{2008AJ....135..527K,2012MNRAS.419.1051L}.
Instead, the distribution of \ion{H}{1} is elongated along the optical minor axis (so-called ``\ion{H}{1} plume''), which is almost at the same direction as the dust lane.
Previous \ion{H}{1} observations revealed that there is a velocity gradient along the dust lane, which was interpreted as cold \ion{H}{1} falling into the galaxy, rather than gas rotation along the minor axis like a polar-ring galaxy \citep{2012MNRAS.419.1051L}. The velocity structure of CO emission also show a similar trend and supports this scenario \citep{2002AJ....124..877M,2015PASJ...67L...1M}.
The origin of the burst in NGC\,5253 is believed to be due to gas infall, which may have been caused by the past interaction with the spiral galaxy M\,83 or other subgroup members \citep{1999AJ....118..797C,2008AJ....135..527K}, and probably started more than 100\,Myr ago \citep{2012MNRAS.419.1051L}.

The main goal of this paper is to study the gas distribution and cloud properties in NGC\,5253 with parsec-scale resolution to be able to resolve clouds close to the SB and to gain insight into the origin of the triggering of an SB event in BCD galaxies.
The outline of this paper is as follows. The observations and
data reduction are summarized in Section\,\ref{obs}. In Section\,\ref{result} we
present the overall molecular gas distribution, the identification of cloud properties, the scaling relations, and the comparison with other galaxies.
In Section\,\ref{discussion}.1, we cross-match the molecular clouds with a compilation of young star clusters from the literature, and we compare differences between the properties of the star forming clouds.
In Section\,\ref{discussion}.2, the overall molecular gas distribution and kinematics are studied and compared with numerical simulations to shed light on the origin of the SB in NGC\,5253.

In a forthcoming paper, we will report the star formation activities around the central SB using H30$\alpha$ emission as well as 230\,GHz continuum data (Paper\,{\sc ii}, Miura, R. E. et al. 2018, in preparation), which were obtained with the same observing setup.
A comparison of the star formation rate (SFR) derived from H30$\alpha$ emission with those from other tracers have already appeared in a separate paper \citep{2017MNRAS.472.1239B}.
A spatial comparison between the 230\,GHz continuum and the CO molecular gas distribution for each cloud will be presented in a separate paper (Paper\,{\sc iii}, Miura, R. E. et al. 2018, in preparation).

\section{Observations}\label{obs}

\begin{figure}
\begin{center}
\includegraphics[width=0.5\textwidth,trim={36 25 20 30}, clip]{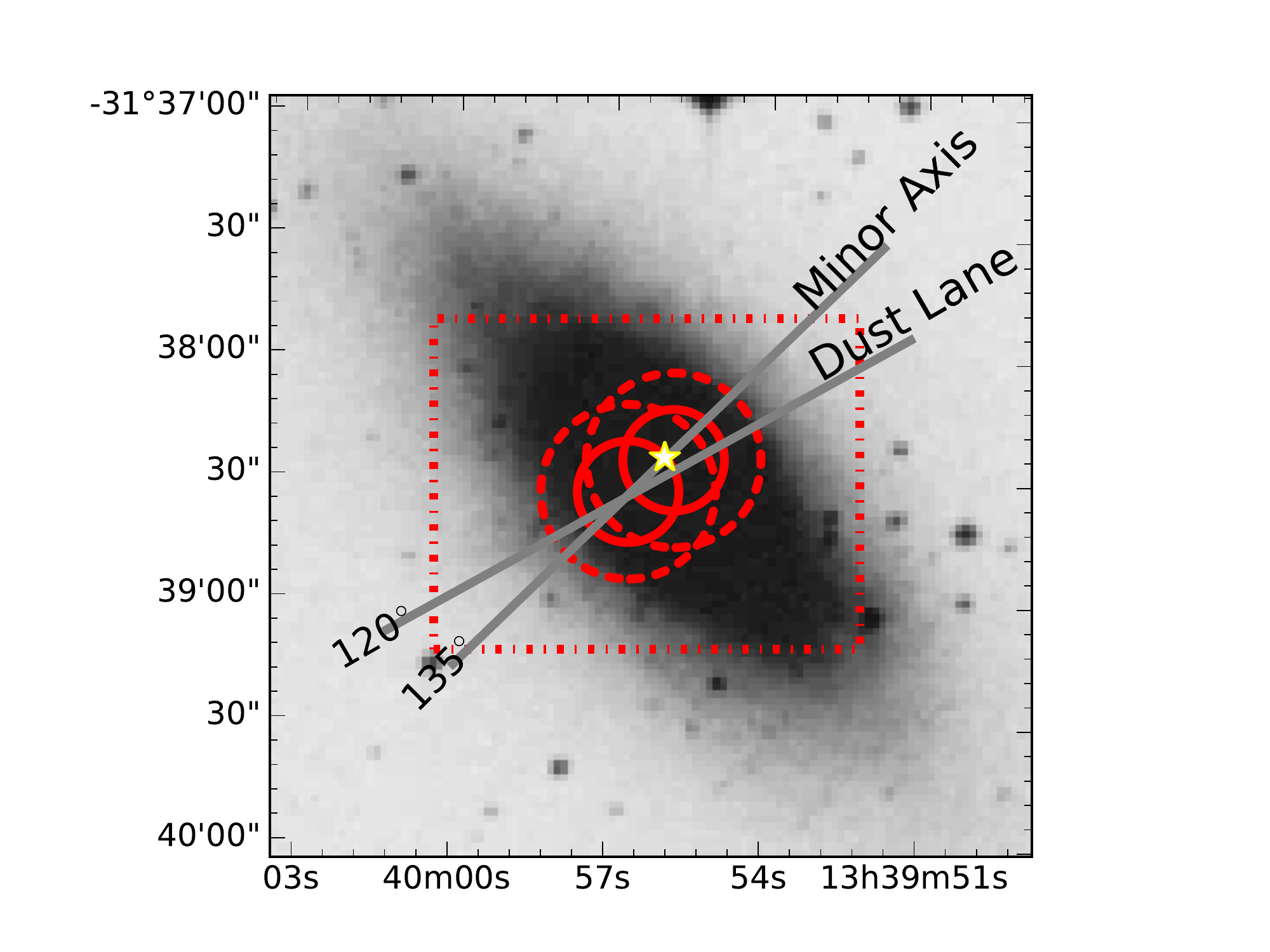}
\caption{
Digital Sky Survey (DSS) optical image with ALMA observational fields. The field of views of the ALMA observations are shown as solid-line circles, dashed-line circles and dot-dashed rectangle for the 12 m, 7 m and TP array, respectively. 
The star symbol corresponds to the location of two massive and young super star clusters. 
The position of the dust lane is indicated. The dust lane is almost along the optical minor axis.
\label{almafov}}
\end{center}
\end{figure}

The ALMA data were obtained during  2014 June and 2015 July using Band\,6 (project code: 2013.1.00210.S; PI: R. E. Miura).
The data consisted of twelve executions: two ACA 7 m array datasets, three 12 m array data sets and seven Total Power (TP) array datasets. The execution IDs and their array configurations are given in Table\,\ref{tbl1}.
We simultaneously observed  $^{12}$CO(2--1) and H30$\alpha$ emissions toward two overlapping pointings covering the dust lane and the center of NGC\,5253 (Figure\,\ref{almafov}). 
The two pointings were centered at $\alpha=13\hr39\mn55\fs87$, $\delta=-31\arcdeg38\arcmin25\farcs9$ and $\alpha=13^{\rm h}39^{\rm m}56\fs77$, $\delta=-31\arcdeg38\arcmin32\farcs4$ (ICRS).
The field of view (half-power beam width, HPBW) is $25\arcsec$ and $43\arcsec$ (at 230\,GHz) for the 12 m and 7 m antennas, respectively.
The raster map of the TP observations was centered at $\alpha=13^{\rm h}38^{\rm m}56\fs3$, $\delta=-31\arcdeg39\arcmin29\farcs2$ with a rectangular field $84\arcsec\times 84\arcsec$.  
The theoretical beam size of the TP image is 28$^{\arcsec}$ at 230\,GHz taking into account the scanning pattern and the grid function used, that is, spheroidal. 

CO(2--1) ($\nu_{\rm rest}$=230.538\,GHz) was observed simultaneously with H30$\alpha$ ($\nu_{\rm rest}$=231.900\,GHz) in the lower side band
and CS(5--4) ($\nu_{\rm rest}$=244.936\,GHz) in the upper side band. 
These three lines were observed in different spectral windows. Each spectral window has a bandwidth of 1875\,MHz with a resolution of 1.129\,MHz (equivalent to 1.470\,\kms\, at the frequency of the CO(2--1) line).
The execution time was about 1 hr 3 minutes for the extended 12 m array, 23 minutes for the compact 12 m array,  2 hr 15 minutes for the 7 m array, and 4 hr 46 minutes for the TP array observation.
The time on source for each execution is listed in Table\,\ref{tbl1}.

% reduction
Data calibration and imaging were manually done using the Common Astronomy Software Applications \citep[CASA;][]{2007ASPC..376..127M} version 4.7.0 for all of the interferometric datasets.
We used bandpass, phase, and flux calibrators as intended for all executions except uid://A002/X8440e0/X29c6. For this data set, we used the bandpass calibrator J1427-4206 also as flux calibrator, instead of using Ceres, because it is known that the model for Ceres in that version of CASA has large offsets from a more reliable flux model constructed from {\it Herschel} data.
All executions have flux calibration uncertainty of less than 10\,\%.

For the TP array data, we performed calibration and imaging using the CASA 4.7.0 Pipeline.
We derived a conversion factor from Kelvin (K) to Jansky (Jy) using the command in {\sf analysisUtils} {\sf getJyPerK} which determines the conversion factor from the empirical relations as a function of antenna, frequency, temperature and elevation.
For the spectral window in which the CO(2--1) line falls, the obtained K-to-Jy conversion factor ranged from 43.9 to 44.4\,Jy/K.
The uncertainty of the conversion factor is expected to be 10\,\% in Band\,6.
Application of the conversion from K to Jy is done inside Pipeline, so that the generated images have units of Jy\,beam$^{-1}$.
With the TP observations, the CO(2--1) data have a rms level of 0.05\,Jy\,beam$^{-1}$ for a 1.5\,\kms\,resolution.
The total flux obtained in the observed area is $96.6\pm0.2$\,Jy\,\kms\,between 352 to 461\,\kms\,and within a box of $90\arcsec\times84\arcsec$.

 % properties of final data cube
We calibrated each of the five interferometric data sets independently and concatenated them after
performing continuum subtraction.
We excluded the CO(2--1) and H30$\alpha$ lines, as well as the band edges, for continuum subtraction.
We confirmed that it was not necessary to correct weights in the {\it u--v} data for CASA versions later than 4.3, as long as the option {\sf calwt=True} was set when applying the calibration tables.
The relative weights of the visibilities in the compact 12 m array data (C34-2/1, longest baseline of 349\,m) are $\sim$1.0,  $\sim1.8$ for the extended 12 m array data (C34-7/6, longest baseline of 1.6\,km), and $\sim$0.3 for the 7 m array data.
These weights were calculated inside CASA by taking into account the system temperature (Tsys) variance, integration time, and antenna diameter.
We generated a CO(2--1) data cube limiting the velocity range to $V_{\rm LSRK}=345$\,\kms--465\,\kms\, in the LSRK frame \footnote{Throughout this paper, the velocities are expressed in the radio convention and the kinematic local standard of rest (LSRK) frame.} and with 1.5\,\kms\ resolution using {\sf clean} task in CASA. 
In the {\sf clean} task, we used Briggs weighting with the parameter robust=0.5.
The pixel size of the image was set to about one-fourth of the synthesized beam, 0\farcs05.

Finally, the mosaicked $^{12}$CO(2--1) interferometric 12 m and 7 m data cube was combined with the TP image using the feathering algorithm.
The integrated flux of the combined image within a 22$\arcsec$ radius is $92.6\pm0.1$\,Jy\,\kms. 
The combined CO image has a typical noise level of 1.3\,mJy\,beam$^{-1}$ for a velocity resolution of 1.5\,\kms\,resolution.
The spatial resolution of the final images is $0\farcs2$ (corresponding to 3.1\,pc at a distance of 3.15\,Mpc).

The final H30$\alpha$ and CS(5--4) images have rms levels of 0.8\,mJy\,beam$^{-1}$ for a 5.0\,\kms\, channel and 2.2\,mJy\,beam$^{-1}$ for a 3.0\,\kms\ channel, respectively.
Note that no CS emission was detected at this sensitivity, even when binning more channels to increase signal-to-noise ratio.
The results on the H30$\alpha$ and continuum data will be reported in separate papers \citep[Miura, R.E. et al. 2018, in preparation; see also][]{2017MNRAS.472.1239B}.

\section{Results}
\label{result}

\begin{figure}
\includegraphics[width=.5\textwidth,trim={10 30 20 40}, clip]{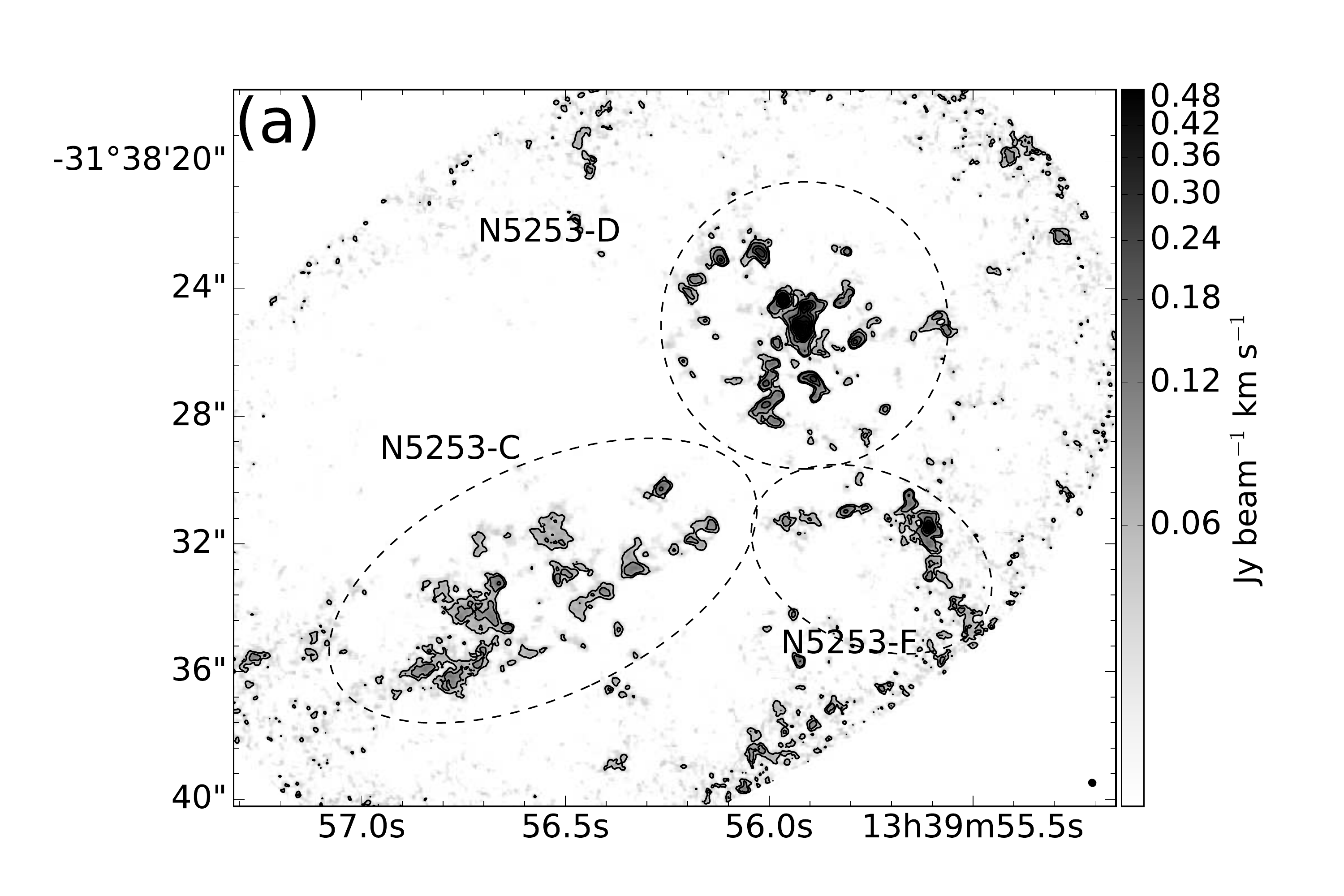}\\
\includegraphics[width=.5\textwidth,trim={10 30 20 40}, clip]{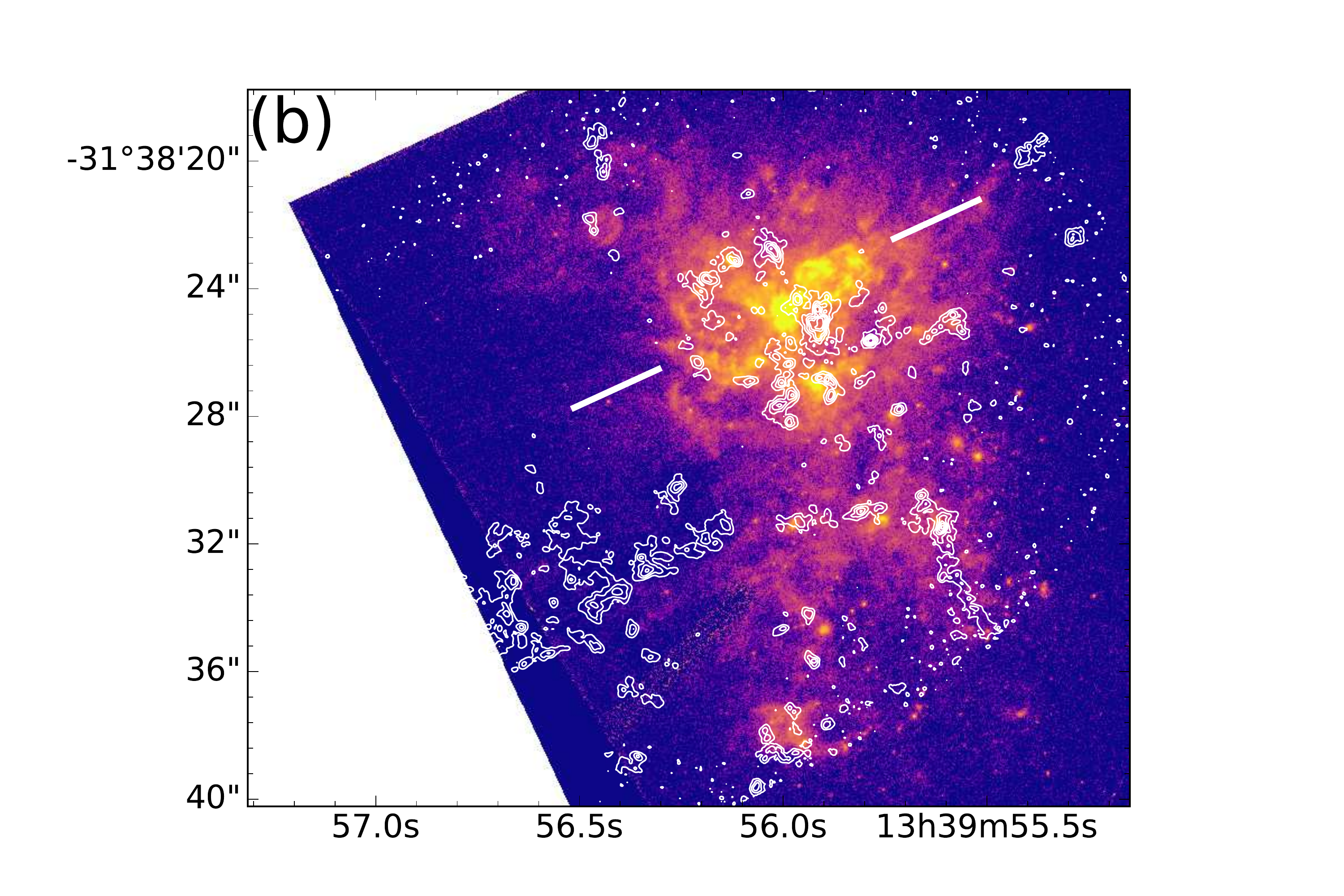}
\caption{(a) CO(2--1) integrated intensity map of NGC\,5253 in color scale and contours. Note that the image is presented in logarithmic scale to show the extended diffuse emission. The contour levels are $3, 5, 10, 15, 20, 25$ and $30\,\sigma$, where $\sigma=0.5$\,Jy\,beam$^{-1}$\kms. 
 The previously identified molecular cloud complexes, N5253-C and N5253-D \citep{2015PASJ...67L...1M,2002AJ....124..877M} are labeled, as well as the newly identified molecular cloud complex N5253-F in this paper. Ellipses indicate the locations of the three molecular cloud complexes.
The synthesized beam is shown as a (black) ellipse at the right bottom corner.
(b) {\it HST}/ACS F658N (H$\alpha$) image on top of the CO(2--1) peak intensity contour map. The color scale is in logarithmic scale. The contour levels are $5, 10, 15, 20, 25$ and $30\,\sigma$, where $\sigma=1.3$\,mJy\,beam$^{-1}$. 
The white bars indicate the location of the H$\alpha$ bipolar outflow.
\label{mom}}
\end{figure}

We are able to resolve molecular cloud complexes into parsec-scale molecular clouds thanks to ALMA's high resolution and sensitivity and high dynamic range.
Figure\,\ref{mom}a shows the CO(2--1) integrated intensity map.
When generating the integrated intensity map shown in the figure, a mask is applied that excludes pixels with low signal-to-noise ratios. The mask only includes pixels whose intensities are over $4\,\sigma$ and adjacent ones down to $2\,\sigma$ level.
The CO emission is distributed mainly in three distinct regions: the SB region, the dust lane, and the region to the SW of the dust lane.
The molecular cloud complexes in the SB and dust lane regions were identified in previous CO(2--1) observations \citep[hereafter N5253-D and N5253-C, respectively, following the nomenclature in][]{2002AJ....124..877M,2015PASJ...67L...1M}.
Note that \citet{2002AJ....124..877M} studied the dust lane in NGC\,5253 with the Owens Valley Millimeter Array, with a resolution of $10\arcsec\times5\arcsec$, while \citet{2015PASJ...67L...1M} observed with the Submillimeter Array at a resolution of $11\arcsec\times4\arcsec$.
This is the first time that the molecular cloud complex to the SW of the dust lane has been detected.
This molecular cloud complex is a spatially and kinematically distinct component with respect to the other two molecular cloud complexes.
Hereafter we call this molecular cloud complex N5253-F, to avoid confusion with the nomenclature used in \citet{2002AJ....124..877M}.

N5253-D is the region where molecular gas is most dominant and where the highest concentration is found.
N5253-C is a more elongated and filamentary cloud distributed along the dust lane and composed of both diffuse and small clouds.
N5253-F is an L-shaped filamentary structure and also a group of several clumpy structures.
About 70\,\% of the total CO(2--1) emission is concentrated in the central $8\arcsec$ (or 122\,pc) of the galaxy (SB region, or N5253-D),
while the rest is extended along the dust lane and the SW region, in agreement with previous observations \citep{2015PASJ...67L...1M,2002AJ....124..877M}. 

\begin{figure}
\includegraphics[width=.5\textwidth,trim={10 30 20 40}, clip]{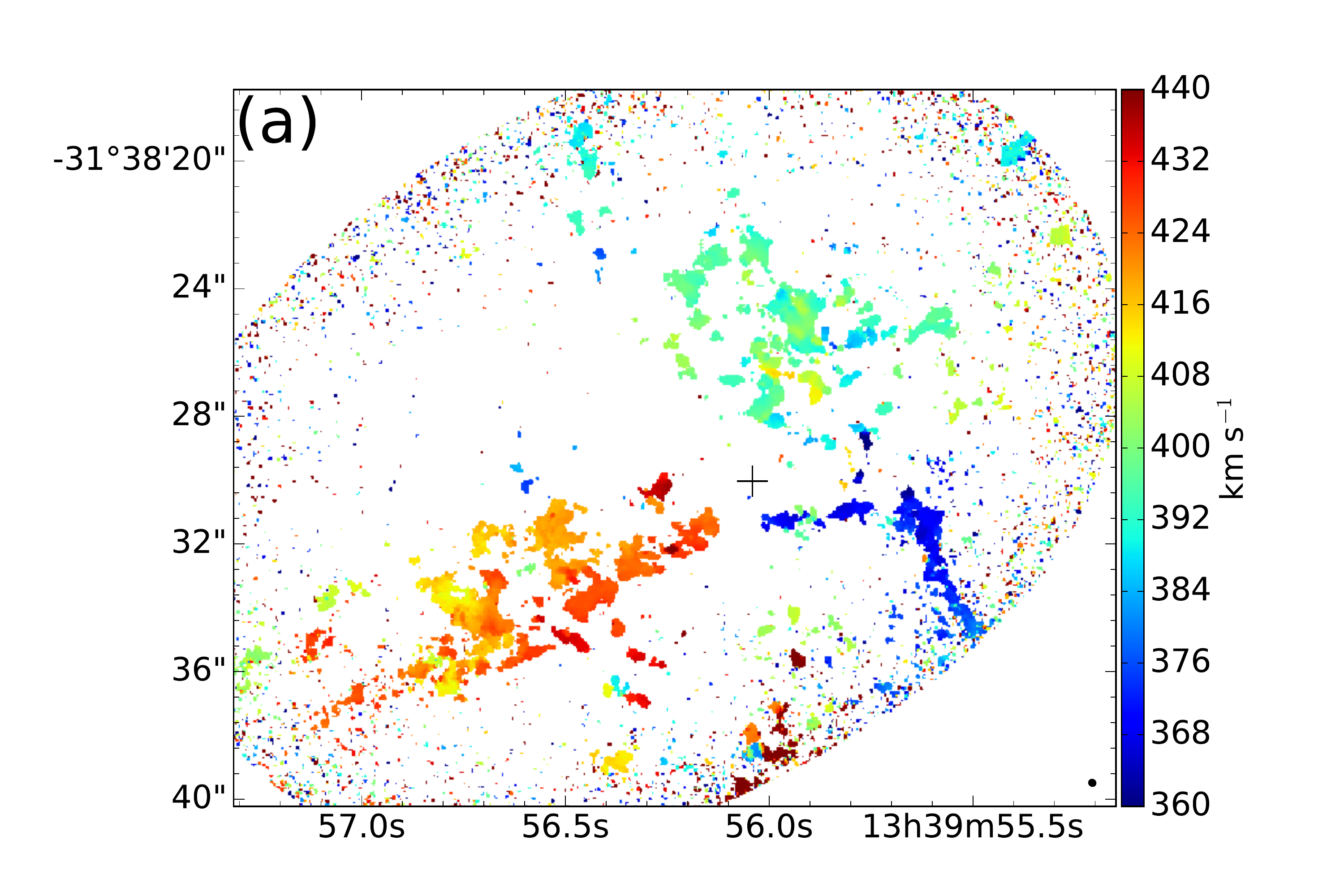}\\
\includegraphics[width=.5\textwidth,trim={10 30 20 40}, clip]{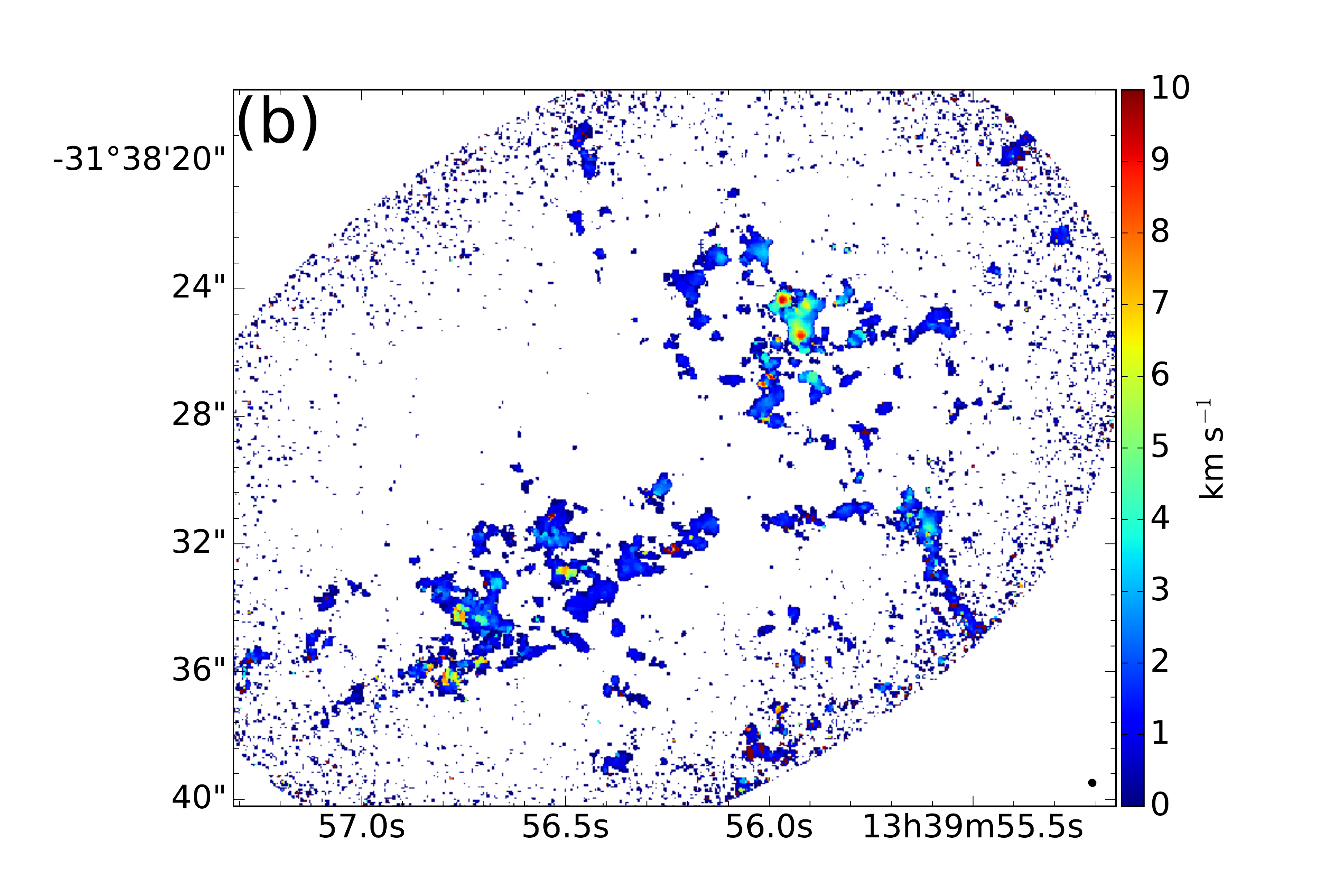}
\caption{(a) Intensity-weighted velocity field in color scale.
The cross indicates the optical center. (b) Velocity dispersion map in color scale.  \label{mom2}}
\end{figure}

Figure\,\ref{mom}b indicates the locations of the CO(2--1) distributions over an {\it HST} H$\alpha$ image.
The dust lane is not fully covered by the field of view of the {\it HST} image.
The SB region, which is very bright in H$\alpha$ emission, is located slightly north of the optical center of the galaxy (see Figure\,\ref{mom2}).
The bipolar outflow is seen across both E-W sides of the H$\alpha$ emission peak \citep[indicated in white-colored bars in Figure\,\ref{mom}b;][]{2013A&A...550A..88W}.
There are also multiple bubbles or filamentary structures around these regions \citep{1997AJ....114.1834C}.
We found that the CO emission is often (except in the dust lane) found around relatively bright H$\alpha$ emission such as in the central SB, and the clumps are found scattered to the south of the SB region.
In the central SB region (N5253-D), the CO emission is distributed mainly along the N-S direction, also avoiding the axis of the H$\alpha$ bipolar outflow.

Figure\,\ref{mom2}a shows the intensity-weighted velocity map.
We do not find a clear sign that the three molecular cloud complexes are following the galaxy rotation.
The three molecular cloud complexes have each have distinct velocity ranges, which are 390--410\,\kms\, for N5253-D, 410--450\,\kms\, for N5253-C, and 360--380\,\kms\, for N5253-F.
The N5253-D component, associated with the central SB region, is close to the systemic velocity of the central SSCs, 389\,\kms\, \citep[based on stellar absorption lines;][]{2004ApJ...610..201S}, while N5253-F and N5253-C have overall approaching and receding velocities from it.

Figure\,\ref{mom2}b shows the velocity dispersion map of NGC\,5253. 
The majority of the molecular clouds have velocity dispersions of a few \kms, except in the central region, N5253-D, where clouds show relatively large velocity dispersions of $\sigma_v>5$\,\kms. 
The largest velocity dispersion $\sigma_v=9.2$\,\kms\, is found at $\alpha=13^{\rm h}39^{\rm m}55\fs97$, $\delta=-31\arcdeg38\arcmin24\farcs37$, at the same position of the highest integrated intensity peak. The second largest velocity dispersion, $\sigma_v=8.6$\,\kms\, is found 1\farcs3 (20\,pc) southwards from it, at $\alpha=13^{\rm h}39^{\rm m}55\fs92$, $\delta=-31\arcdeg38\arcmin25\farcs47$.

% CO(2-1) moment map

\subsection{Position-Velocity Diagrams} \label{filament}
\begin{figure}
\includegraphics[width=0.5\textwidth]{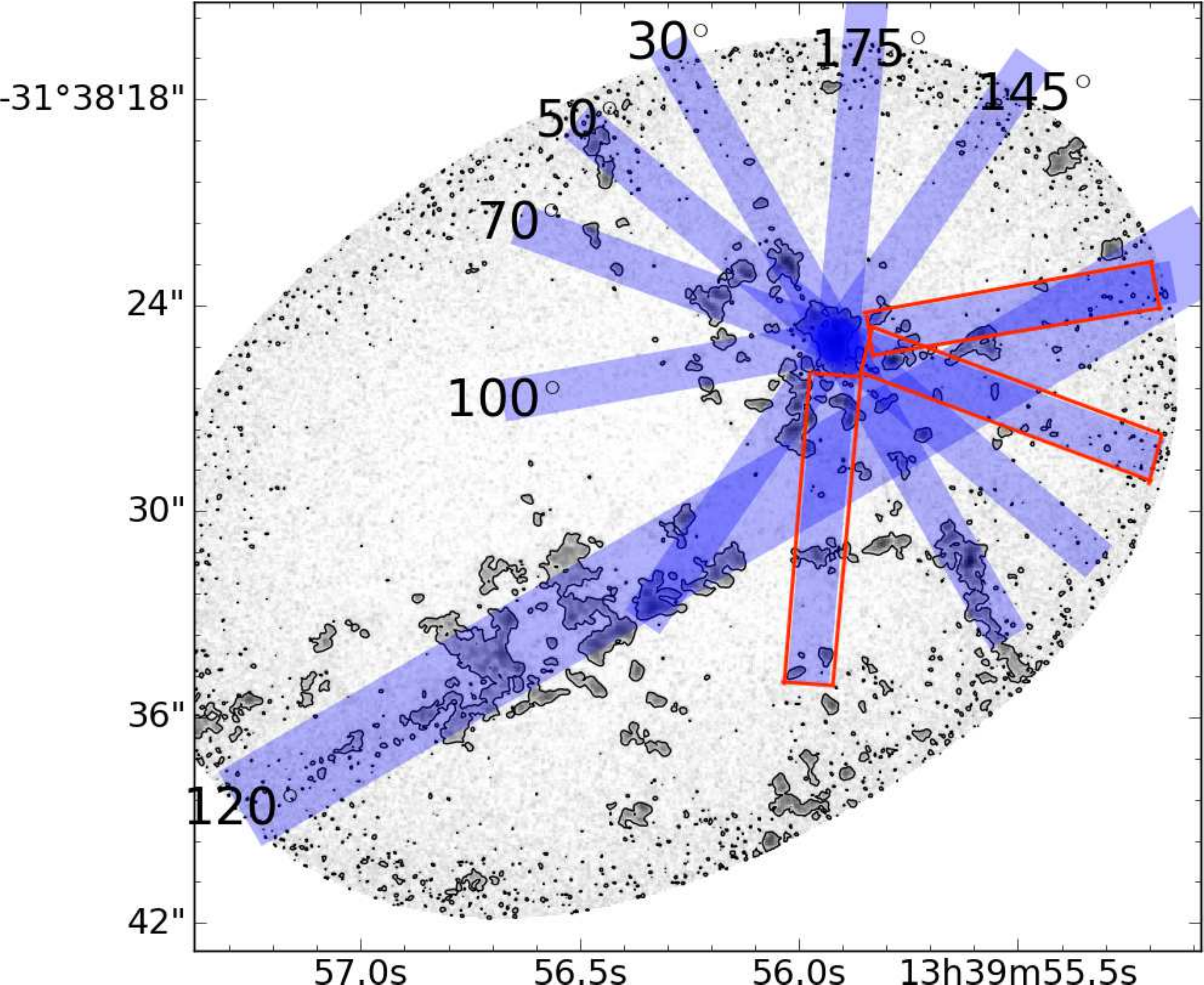}
\caption{The Position-Velocity cuts used in Figure\,\ref{pvcut}, over the CO(2--1) peak intensity map. The contour is $5\,\sigma$.
The red boxes correspond to the regions where there are structures with smooth velocity gradients in the PVDs and not following galaxy rotation. These components are also indicated in Figure\,\ref{pvcut}. }\label{mom8}
\end{figure}

\begin{figure*}
\includegraphics[width=.5\textwidth,trim={0 0 0 0}, clip]{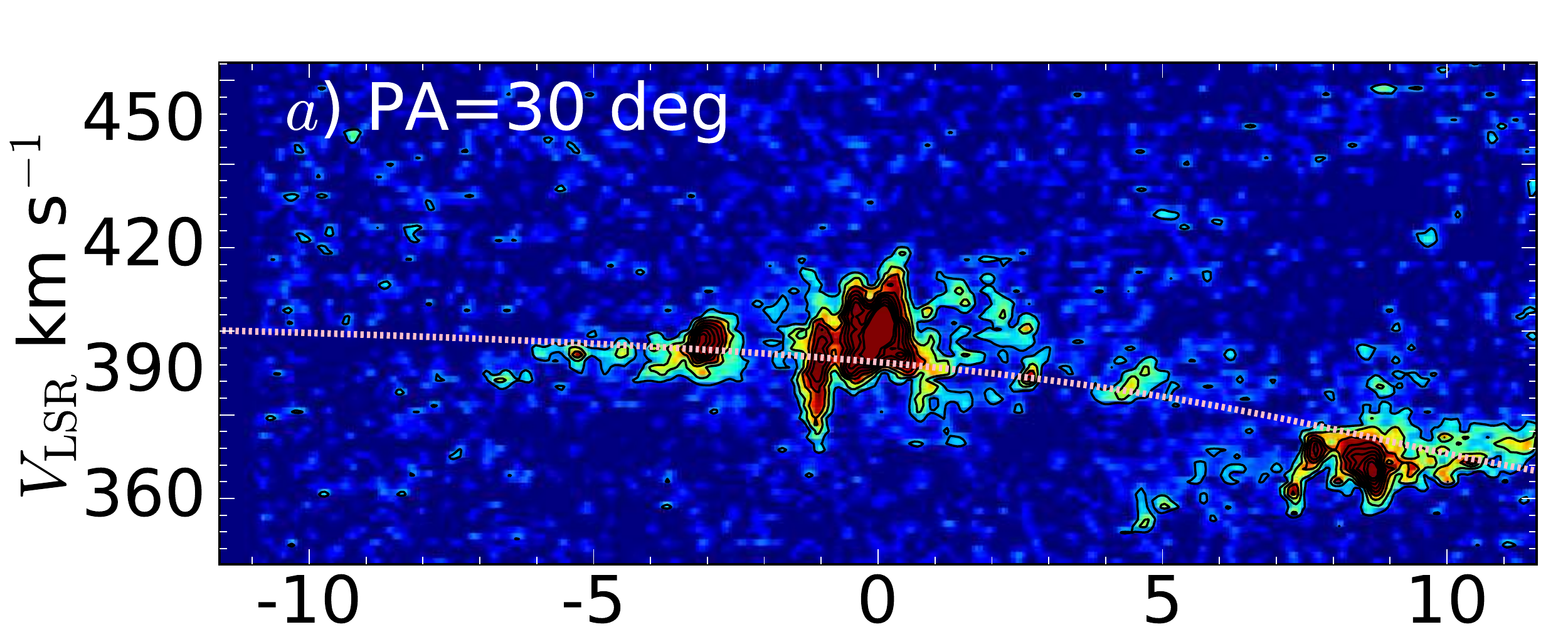}
\includegraphics[width=.5\textwidth]{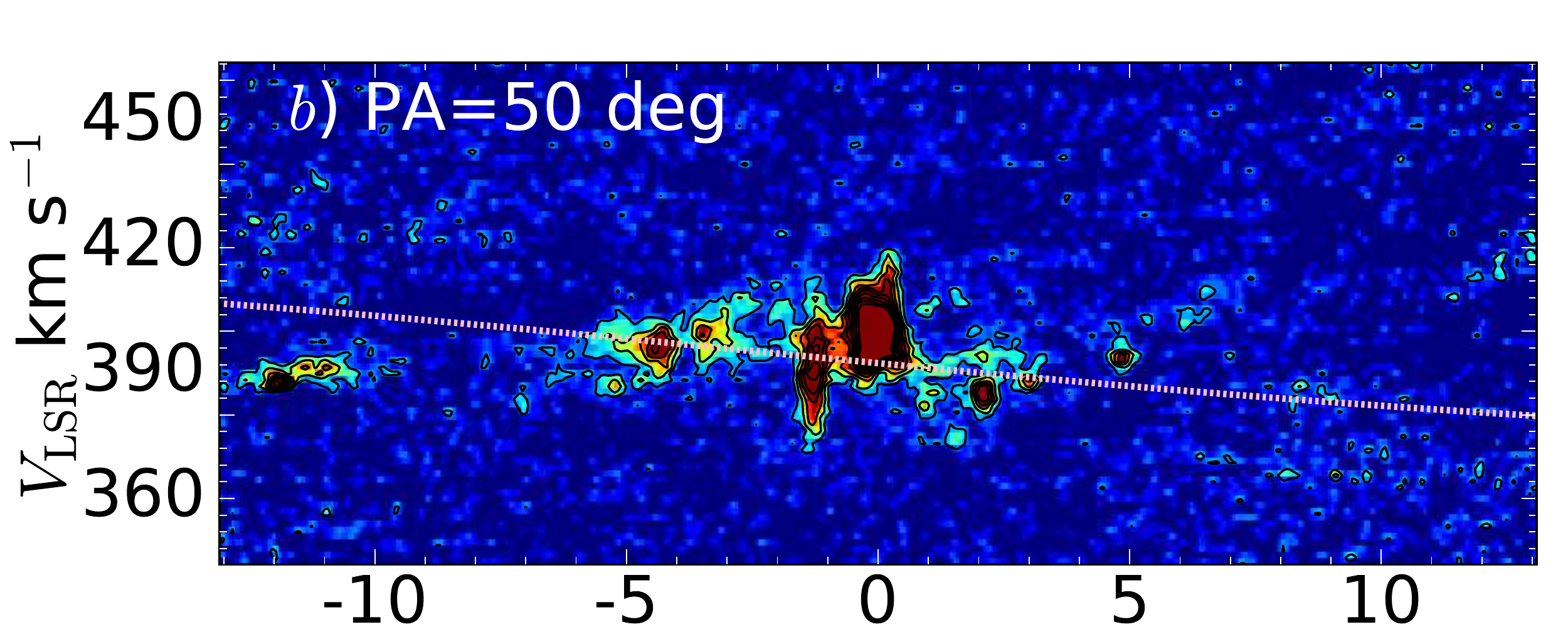}
\includegraphics[width=.5\textwidth]{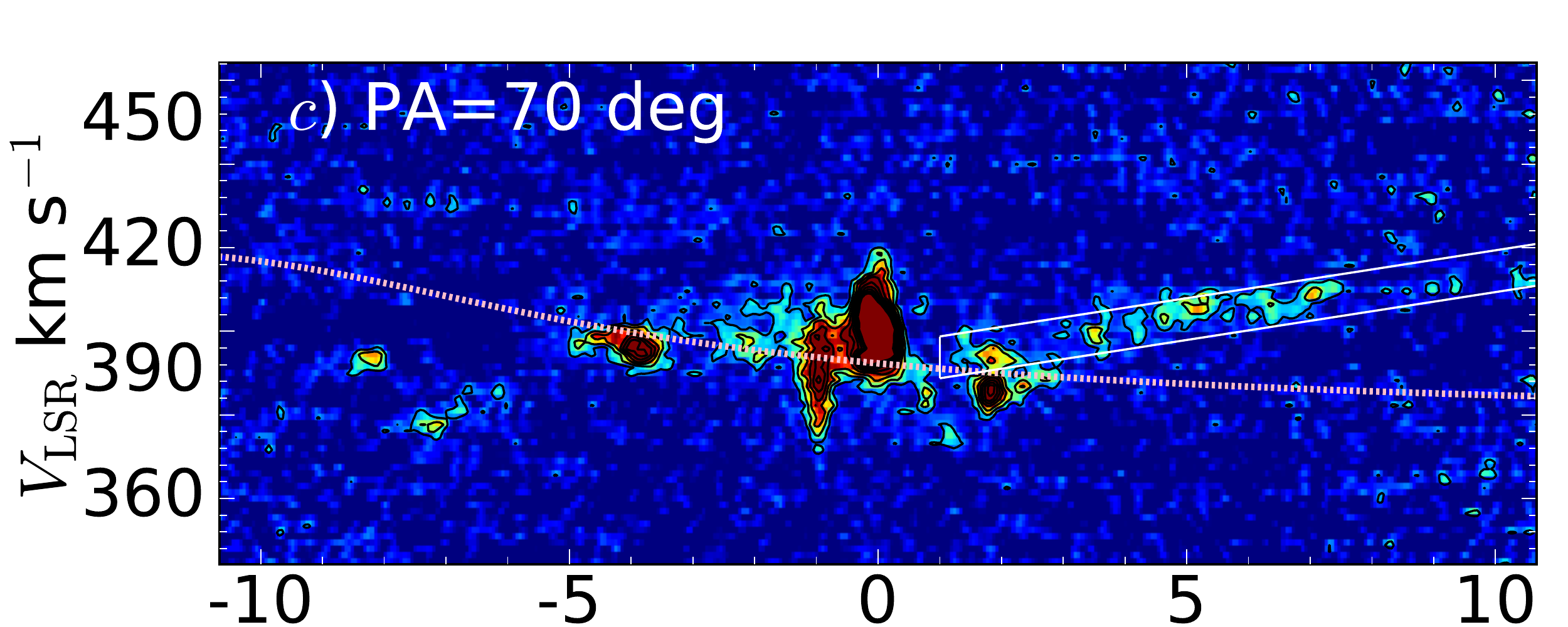}
\includegraphics[width=.5\textwidth]{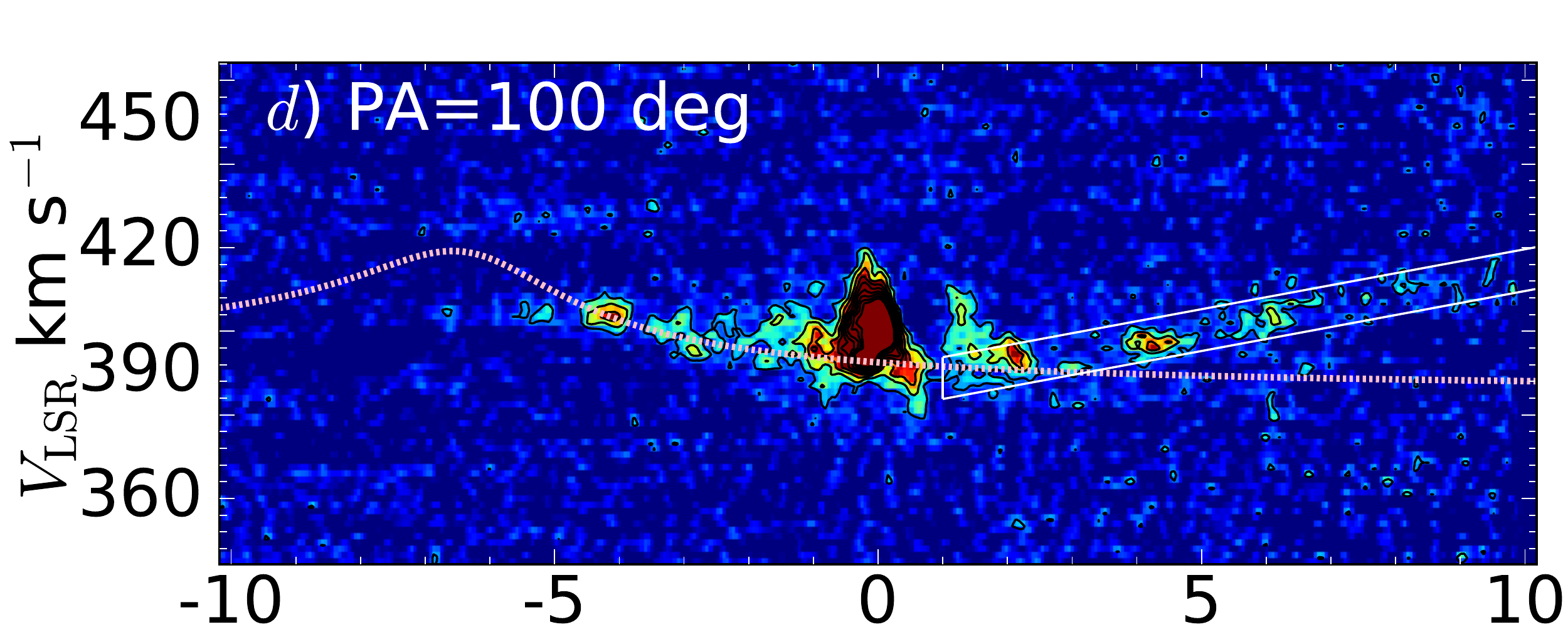}
\includegraphics[width=.5\textwidth]{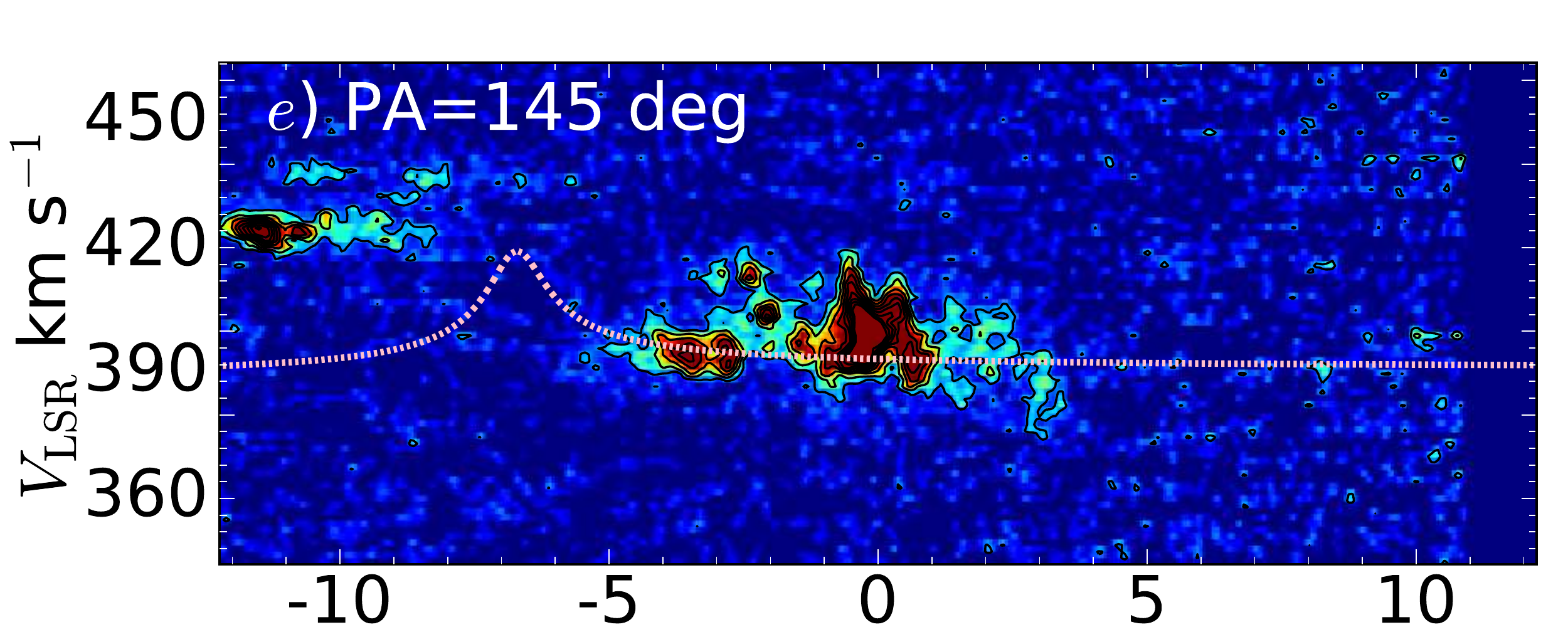}
\includegraphics[width=.5\textwidth]{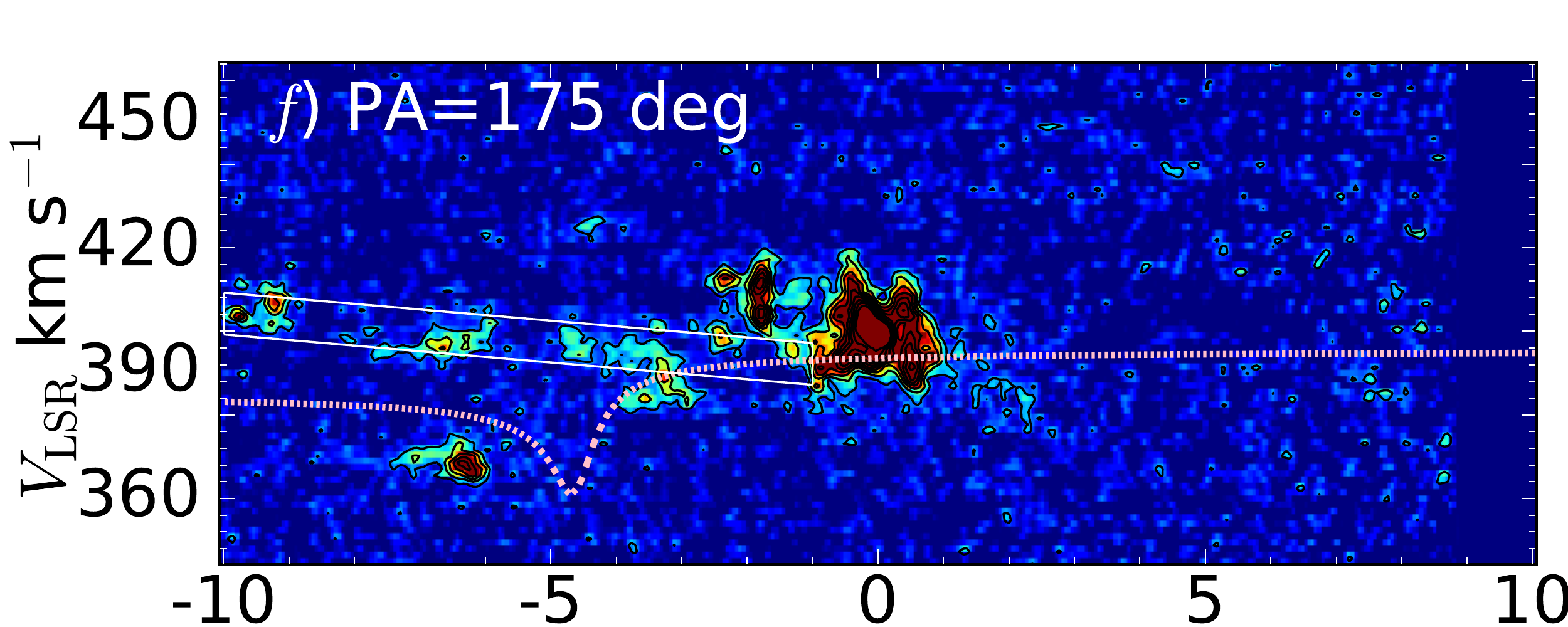}
\includegraphics[width=.8\textwidth]{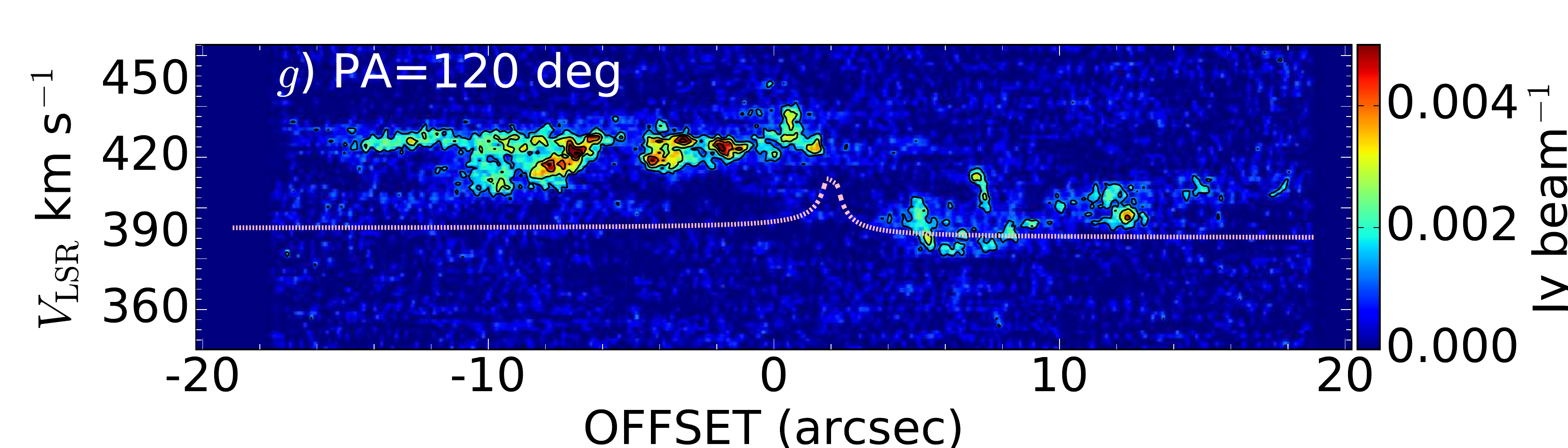}
\caption{
The CO(2--1) position-velocity diagrams, as indicated in Figure\,\ref{mom8}. The x axis is shown as relative coordinates in arcsec from $(\alpha,\delta)=(13^{\rm h}39^{\rm m}55\farcs911,-31\arcdeg38\arcmin25\farcs044)$ for $(a)$--$(f)$ and $(\alpha,\delta)=(13^{\rm h}39^{\rm m}56\farcs266,-31\arcdeg38\arcmin31\farcs174)$ for $(g)$.  Positive in right ascension corresponds to left in the horizontal axis. The color scale is shown in the last panel, and the unit is Jy\,beam$^{-1}$.
The dashed lines indicate the expected galaxy rotation if the rotation velocity is 30\,\kms. 
The white boxes indicate the structures with smooth velocity gradients in the PVDs and not following galaxy rotation.
\label{pvcut}}
\end{figure*}

Figure\,\ref{mom8} and \ref{pvcut} show the position-velocity diagrams (PVDs) and cuts with position angles PA = $30\arcdeg, 50\arcdeg, 70\arcdeg, 100\arcdeg, 145\arcdeg$ and $175\arcdeg$, centered at the CO(2--1) emission peak, and along the dust lane (PA = $120\arcdeg$).
The cut width is 1$\arcsec$ for all diagrams except the one along the dust lane, which is $1\farcs75$ to cover the whole width of the dust lane.
The center of N5253-D is located at (0$\arcsec$, 400\,\kms) in the PVDs at PA = $30\arcdeg, 50\arcdeg, 70\arcdeg, 100\arcdeg, 145\arcdeg$ and $175\arcdeg$.
The PVD at PA = $30\arcdeg$ is a cut along the centers of N5253-D and N5253-F.
The center of N5253-F is located at around (8$\arcsec$, 370\,\kms).
 As seen in the integrated intensity map, N5253-F has a filamentary structure toward the SW, which is a coherent structure as seen in the PVD, that is, the velocity is gradually blue-shifted as the gas moves from S to N. We do not see a clear connection between N5253-D and N5253-F in the PVDs as well as in the integrated intensity map.

We note that we found a compact and abnormally large velocity width component in N5253-D around ($-1$\farcs$7, 390$\,\kms) in the PVD at PA = $30\arcdeg$.
This coincides with the location and systemic velocity of the H30$\alpha$ emission peak \citep{2017MNRAS.472.1239B}, as well as the central SSCs \citep{2000ApJ...532L.109T,2004ApJ...610..201S}.
The full width at zero intensity (FWZI) is about 25\,\kms\ and the spatial extent is about 0\farcs2 (or 3\,pc), which means that it is an unresolved component.
We will report the peculiarities of this component in Paper\,{\sc ii}.

The PVDs at PA = $50\arcdeg, 70\arcdeg$ and $100\arcdeg$ show that extended emission along the NE-SW direction of N5253-D is linked to its central and main molecular gas component.
The relatively smooth velocity gradient on the PVD is about $-2.3$ to $+2.9$\,\kms\,arcsec$^{-1}$.

The PVD at PA = $145\arcdeg$ also show that the relatively clumpy structures located at ($-3\arcsec, 390$\,\kms) are smoothly connected to the major component of N5253-D.
Note that the emission around ($-11\arcsec, 420$\,\kms) is part of N5253-C.
The two components around ($-10\sim-5\arcsec, 370\sim400\,\kms$) in the PVD at PA = $175\arcdeg$ correspond to emission located between N5253-C and N5253-F in Figure\,\ref{mom8},
and have distinct velocities, one at around 400\,\kms\ and another at about 370\,\kms. The former is again connecting the main component in N5253-D at (0\arcsec, 400\,\kms).

The PVD at PA = $120\arcdeg$ along the dust lane (Figure\,\ref{pvcut}g) shows that there are two main components; that is, one at around $-12\arcsec\sim+1\arcsec$ with an almost constant velocity of 420\,\kms, which belongs to N5253-C, and another around ($+4\sim+12\arcsec, 380\,\kms\sim400$\,\kms), which is part of N5253-D. There is a velocity jump between both components of about 30\,\kms, which confirms the finding of previous lower sensitivity and angular resolution CO maps \citep{2002AJ....124..877M,2015PASJ...67L...1M}.

\subsection{CO(2--1) Molecular Cloud Catalog} \label{clump}

We identified molecular gas clouds using the package CPROPS in IDL and derived their properties.
For details about the algorithm to identify clouds and the definition of the derived cloud properties, please refer to \citet{2006PASP..118..590R}.
Briefly, we summarize how the procedure works. 
We used default values for most of the parameters in the algorithm, and we masked emission below five\,$\sigma$.
The algorithm begins by searching for emission that appears in connected, discrete regions ({\it Islands}), and meanwhile we exclude other regions  ({\it Islands}) that do not have a minimum projected area of one spatial resolution element (synthesized beam $\sim0\farcs2$) and a minimum velocity width of two channels (3.0\,\kms).
The algorithm also performs decomposition on the identified {\it Islands} with more than two spatial resolution elements by searching local maxima for each {\it Island}.
If there is more than one local maximum for a given contour intensity level ({\it merge level}), the emission above the {\it merge level} will be identified as a deconvolved cloud, unless each component shows at least a $2\sigma$ difference between its local maximum and {\it merge level}, and also has more than one spatial resolution element. 
Some cloud candidates identified by this procedure outside the beam width at a 60\,\% power level are excluded so as to avoid larger uncertainties caused by primary beam correction and false detections at the edges of the field of view.
We also excluded clouds with low signal-to-noise ratio of less than 5 and also with the size nearly equivalent to the spatial resolution, because the extrapolation done inside the algorithm would depart systematically from the true value in such a case \citep{2006PASP..118..590R}.

As a result, we have identified a total of 118 molecular clouds, whose properties are summarized in 
Table\,\ref{tbl2}.
The columns indicate
the cloud ID, the cloud position  in relative coordinates to the center ($\Delta\alpha$, $\Delta\delta$, where the center is $\alpha=13^{\rm h}39^{\rm m}56\fs041$, $\delta=-31\arcdeg38\arcmin30\farcs03$), the velocity ($v_{\rm LSR}$), the velocity dispersion ($\sigma_v$),
the radius ($R$), the CO(2--1) flux density ($S_{\rm CO(2-1)}$), and 
virial mass ($M_{\rm VIR}$).
All measurements were calculated by extrapolating the emission profile to the zero intensity level.
The radius is defined as $R=1.91\,\sqrt{[\sigma_{\rm major}^2-\sigma_{\rm beam}^2]^{1/2}[\sigma_{\rm minor}^2-\sigma_{\rm beam}^2]^{1/2}}$, where $\sigma_{\rm major}$ and $\sigma_{\rm minor}$ are the (extrapolated) rms sizes of the major and minor axes of the cloud, $\sigma_{\rm beam}$ is the synthesized beam size, and the coefficient 1.91 converts the rms sizes to effective spherical radius of the cloud using the factor $3.4/\pi^{1/2}$, following \citet{1987ApJ...319..730S}.
In this equation, the cloud size is corrected by the angular resolution bias by subtracting the spatial resolution element.
The velocity dispersion is also corrected by resolution bias using equation $\sigma_v=\sqrt{\sigma_{v,{\rm undeconv}}^2-\frac{\sigma_{v,{\rm chan}}^2}{2\pi}}$, where $\sigma_{v,{\rm undeconv}}$ is the extrapolated velocity dispersion, and $\sigma_{v,{\rm chan}}$ is the velocity resolution element.
We compute the virial mass using $M_{\rm vir}=189\,\Delta V^2\,R$ [\Msol], under the assumption that each cloud is spherical and virialized with a density profile described by a truncated power law of the form $\rho\propto r^{-1}$. 
Here, $\Delta V$ is the FWHM velocity line-width in \kms\,and expressed by $\Delta V = 2\sqrt{2\ln 2}\, \sigma_v$.

The uncertainty of each parameter was assessed using a bootstrapping method employed in CPROPS.
This method randomly generates a trial cloud within an allowed number of points in the cloud, derives the cloud properties of the trial cloud,
and then calculates the median absolute deviation of those properties.
We repeated this resampling and remeasuring step 10,000 times for each cloud, and we estimated the uncertainties. 
The final uncertainty in each property is  the median absolute deviation of the bootstrapped values
scaled up by the square root of the oversampling rate, which is equal to the number of pixels per synthesized beam.
Note that the uncertainty in the table does not include the intrinsic error of the flux nor the spatial and velocity resolution limit in the data cubes.

\begin{figure}
\begin{center}
\includegraphics[width=0.5\textwidth,trim={10 45 20 40}, clip]{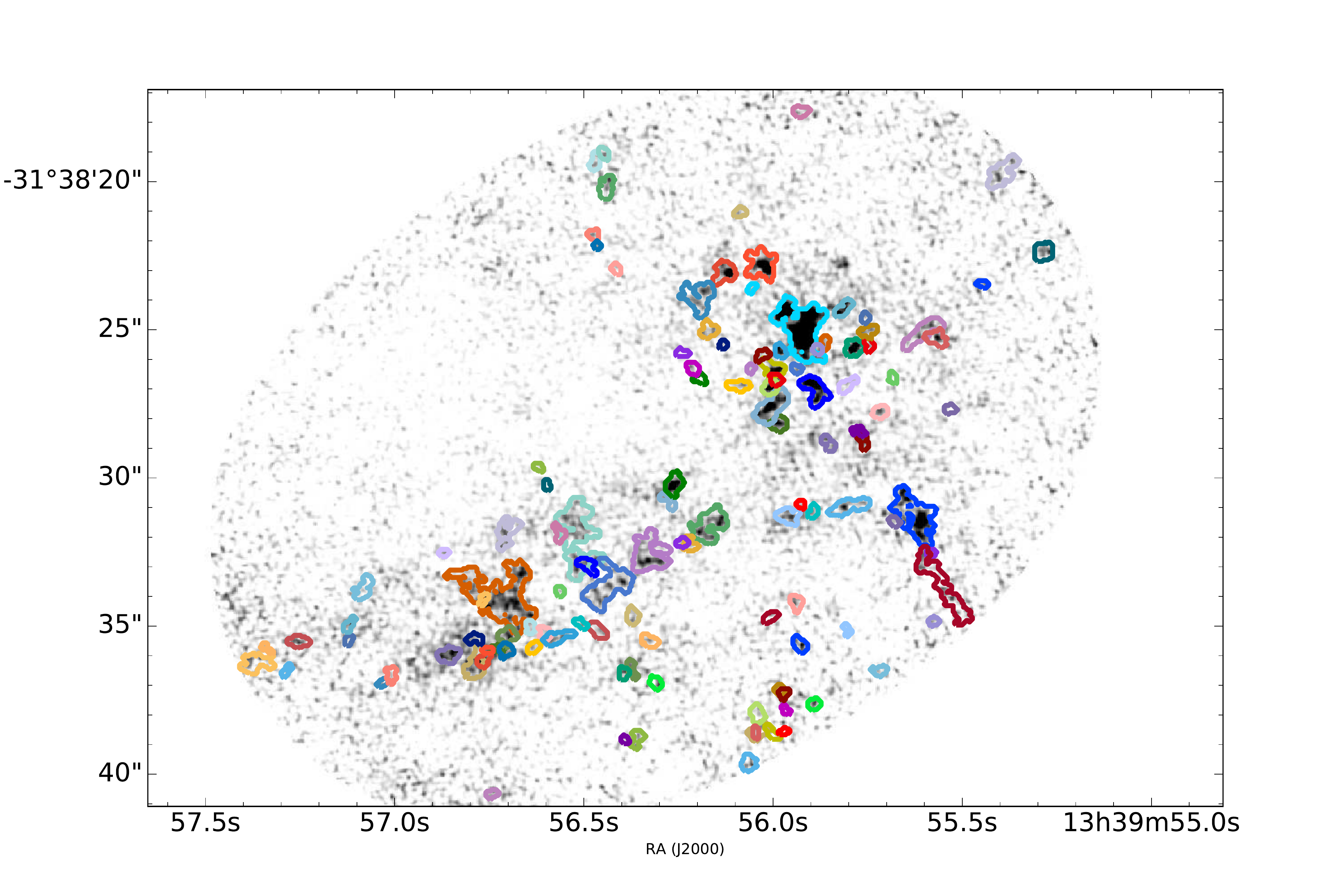}
\end{center}
\caption{The identified molecular clouds in NGC\,5253, overlaid on the CO integrated intensity image. \label{astrodendro}}
\end{figure}

The locations of the identified molecular clouds are shown in Figure\,\ref{astrodendro} overlaid on the integrated intensity image.
Each of the three major molecular cloud complexes, N5253-C, D, and F, are resolved into tens of clustered smaller clouds, 32, 32, and 10 clouds, respectively, within the ellipses indicated in Figure\,\ref{mom}.
The largest molecular cloud among the identified components, cloud\,8 in Table\,\ref{tbl2}, is located in the center of N5253-D.
Cloud\,8 has a clumpy structure inside, and the location of one of the clumps coincides with the central young SSCs.
We will discuss in more detail the cloud properties that are associated with the central SSCs in Paper\,{\sc ii}.
Relatively smaller and more diffuse molecular clouds are also identified in these three molecular cloud complexes.
In the following subsections, we explain the cloud properties and compare with other galaxies, including our Galaxy.

\subsubsection{Line Width - Size Relation} \label{larson-text}
The molecular clouds are known to follow an empirical  line-width - size relation \citep{1981MNRAS.194..809L}.
This relation is a measure of the turbulent conditions of the molecular ISM.
In general, the line-width $\sigma_v$ [\kms] increases as a power of the radius $R$ [pc]: for example, $\sigma_v=0.72\,R^{0.5}$, in the case of our Galaxy \citep{1987ApJ...319..730S,2009ApJ...699.1092H}. 

We find that the line-width and radius of the molecular clouds in NGC\,5253 are in the range of 0.5--8.1\,\kms, and 0.9--20.9\,pc, respectively, with average values of 2.2\,\kms\, and 4.3\,pc. 
The relation of these two parameters is plotted in Figure\,\ref{larson}, and is compared with that of our Galaxy and other nearby galaxies: outer Galaxy \citep{2001ApJ...551..852H}, inner Galaxy \citep{2009ApJ...699.1092H}, the Galactic Center \citep{2001ApJ...562..348O}, Large Magellanic Cloud \citep{2011ApJS..197...16W}, dwarf galaxies \citep{2008ApJ...686..948B}, the SB galaxy NGC\,253 \citep{2015ApJ...801...25L} and other nearby spiral galaxies \citep{2013ApJ...772..107D}.  
All of the samples used in this comparison are identified using CPROPS, and their properties are determined in a similar manner, including the extrapolation in the identification of clouds and the definition of the physical properties \citep[see also][and references therein]{2015ApJ...801...25L}.

The angular resolution of our ALMA data is high enough that we are able to compare with the Galactic molecular clouds.
Generally, quiescent molecular clouds in normal galaxy disks follow the Galactic line-width - size relation,
while those in extreme environments such as our Galactic Center and SB regions follow a similar relation but with an offset of about one dex higher velocity dispersions \citep[e.g.][]{2001ApJ...562..348O}.
 We found that the clouds in NGC\,5253 have on average a slight offset of $\sim$0.2\,dex from the general line-width - size relation found in our Galaxy \citep[dashed line;][]{1987ApJ...319..730S} and other nearby galaxies including dwarf galaxies \citep{2008ApJ...686..948B},  although not as much as in the Galactic Center environment.

In Figure\,\ref{larson} we plot with a different symbol the clouds within the inner $5\arcsec$ (76\,pc) of the center of N5253-D (red circles with a white star inside), so as to highlight the clouds that might be affected by the central SB region.
The clouds near the SB region tend to have larger velocity widths by $\sim$ 0.5\,dex for a given radius than the rest in NGC\,5253, although they are still below the correlation for Galactic Center clouds \citep[][]{2001ApJ...562..348O}.

We do not see any other significant dependency in this plot related to their location or parent molecular cloud complex, such as dust lane (N5253-C) or the southern region of the SB region (N5253-F).
In Figure\,\ref{mom2}b, there are a few locations showing larger velocity widths in the middle of N5253-C and N5253-F. 
The spectra of these locations show double peak features and thus are likely different molecular clouds along the line of sight.

\subsubsection{$I_{\rm CO}-N({\rm H}_2$) conversion factor in NGC\,5253} \label{xcofactor}

We report for the first time the  $I_{\rm CO}-N({\rm H}_2$) conversion factor (hereafter $X_{\rm CO}$ factor) for this relatively metal-poor and SB galaxy.
One of the most popular methods to derive the $X_{\rm CO}$ factor is to compare the CO luminosity and the virial masses of the clouds.
The CO(2--1) luminosity is given by $L^{\prime}_{\rm CO(2-1)} = (c^2/2k_B)\,S_{\rm CO(2-1)}\,\nu_{\rm rest}^{-2}\,D_{\rm L}^2$, or $L^{\prime}_{\rm CO(2-1)} = 3.25\times10^7\,S_{\rm CO(2-1)}\,\nu_{\rm rest}^{-2}\,D_{\rm L}^2$ [K\,\kms\,pc$^2$], where $c$ is the light speed, $k_B$ is the Boltzmann constant, $S_{\rm CO(2-1)}$ is the integrated CO(2--1) line flux density in Jy\,\kms, $\nu_{\rm rest}$ is the rest frequency in GHz, and $D_{\rm L}$ is the luminosity distance to the source in Mpc \citep[][]{2005ARA&A..43..677S}. 
The gas mass of the clouds is calculated as $M_{\rm gas} =  \alpha_{\rm CO}\,L^{\prime}_{\rm CO(1-0)}$, where $\alpha_{\rm CO}$ is a mass-to-light ratio which includes a factor of 1.36 to account for the presence of helium, and $L^{\prime}_{\rm CO(1-0)}$ is the CO(1--0) luminosity \citep{2013ARA&A..51..207B}.
For $X_{\rm CO}=2\times10^{20}$\,cm$^{-2}$(K\,\kms)$^{-1}$, the corresponding $\alpha_{\rm CO}$ is 4.3\,\Msol(K\,\kms\,pc$^{-2}$)$^{-1}$ \citep{2013ARA&A..51..207B}, and thus 
$M_{\rm gas}$ can be expressed as a function of $X_{\rm CO}$ (in units of $10^{20}$\,cm$^{-2}$(K\,\kms)$^{-1}$) as $M_{\rm gas}=4.3\,(X_{\rm CO}/2)\,L^{\prime}_{\rm CO(2-1)}\,R_{2-1/1-0}^{-1}$, where
$R_{2-1/1-0}$ is the CO(2--1) to CO(1--0) integrated intensity ratio.

First, we estimate $R_{2-1/1-0}$ using our ALMA TP CO(2--1) data and previous single-dish CO(1--0) data from the literature.
Single-dish CO(1--0) data are available in \citet{1998AJ....116.2746T} and \cite{1989A&A...225....1W},
in which the NRAO\,12\,m and SEST 15\,m telescopes were used.
Since the half-power beam size (HPBW = 44$\arcsec$) of the SEST telescope matches the field of view of our observations, we chose this measurement for the calculation of the ratio.
We assumed that the observed pointing position was chosen from the {\it IRAS} Catalog of Galaxies and Quasars \citep{1985cgqo.book.....L}, $\alpha=13^{\rm h}39^{\rm m}55\fs33$, $\delta=-31\arcdeg38\arcmin25\farcs23$, which they refer to.
 \citet{1989A&A...225....1W} reported a CO(1--0) integrated intensity of 1.3\,K\,$\kms$ with a velocity width of 43\,\kms.
For the ALMA TP data, we obtained an integrated flux of $42.3$\,Jy\,\kms\, in the velocity range 385--428\,\kms\ and within a 44$\arcsec$ beam area, corresponding to an integrated intensity \footnote{ 
The flux density was converted to brightness temperature using $T_{b}=\lambda^2 S/(2 k_B \Omega_{\rm beam})$, where $\lambda$ is the observed frequency, $\Omega_{\rm beam}$ the beam size, $S$ the flux density, and $k_B$ Boltzmann constant.} of 1.25\,K\,\kms .
We then obtain $R_{2-1/1-0}$ of 0.96 with an uncertainty of $\sim20$\,\%, which is in good agreement with the previous estimate $R_{2-1/1-0}=0.93$ found by \citet{2001AJ....121..740M}.

In our Galaxy, $R_{2-1/1-0}$ is relatively low (0.4--0.6) in the peripheries of molecular cloud complexes, while high (0.7--1.0) in the center of molecular cloud complexes and very high ($>1.0$) toward the center of \ion{H}{2} regions \citep{1994ApJ...425..641S,1997ApJ...486..276S,2001ApJS..136..189S,2015ApJS..216...18N}. 
In SB systems, $R_{2-1/1-0}$ is usually reported to be over unity \citep{2000ApJ...531..200M,2008ApJ...683...70H,2001A&A...365..571W}.
The obtained $R_{2-1/1-0}$ in NGC\,5253 is the average ratio over clouds with different star formation activities, since the 44$\arcsec$ beam ($\sim$670\,pc) includes both the SB region and relatively quiescent clouds compared to the former.
We use unity for $R_{2-1/1-0}$ throughout this paper, but with the caveat that the ratio can change considerably from cloud to cloud depending on the environment.

\begin{figure}
\includegraphics[width=0.5\textwidth]{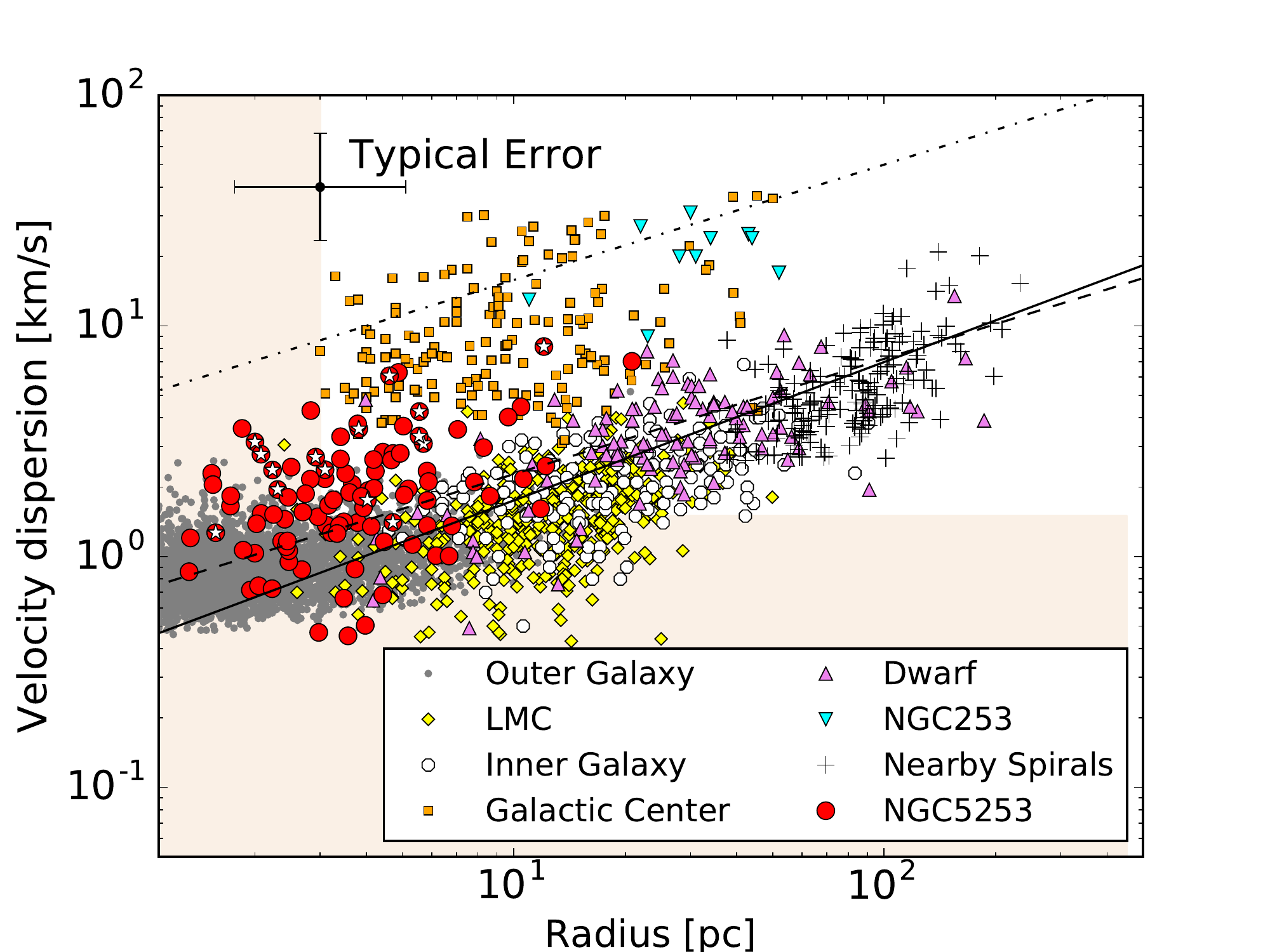}
\caption{
Line width-size relation of the identified clouds in NGC\,5253 (red circles, highlighting the SB region with white star symbols inside the red circles), compared to the (giant) molecular clouds found in the outer Galaxy \citep{2001ApJ...551..852H}, inner Galaxy \citep{2009ApJ...699.1092H}, Large Magellanic Cloud \citep{2011ApJS..197...16W}, the Galactic Center \citep{2001ApJ...562..348O}, dwarf galaxies \citep{2008ApJ...686..948B}, the SB galaxy NGC\,253 \citep{2015ApJ...801...25L} and other nearby spiral galaxies \citep{2013ApJ...772..107D}.  The dashed, dot-dashed and solid  lines indicate the correlation found in our Galaxy \citep{1987ApJ...319..730S}, in the Galactic Center \citep{2001ApJ...562..348O}, and in other galaxies \citep{2008ApJ...686..948B}, respectively. 
The shaded areas indicate the resolution limits obtained from the synthesized beam and the velocity resolution.
\label{larson}}
\end{figure}

\begin{figure}
\includegraphics[width=0.5\textwidth]{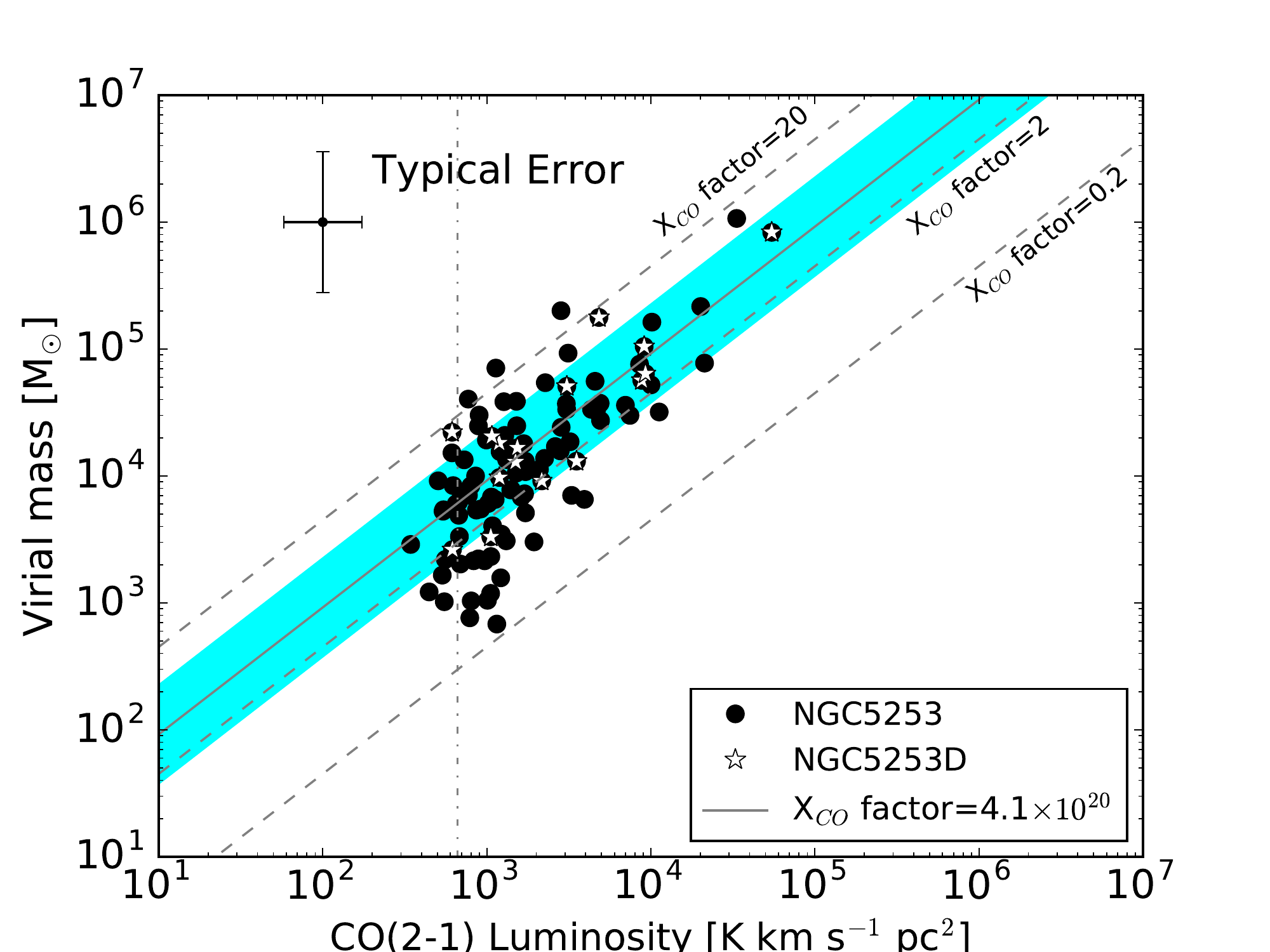}
\caption{
CO luminosity ($L^{\prime}_{\rm CO(2-1)}$) - virial mass relation for the molecular clouds in NGC\,5253. 
The symbols with a star inside indicate the clouds close to the central SB.
The solid line indicates the best linear fit $M_{\rm vir}=(9.2^{+13.2}_{-5.4})\,L^{\prime}_{\rm CO(2-1)}{R_{2-1/1-0}}^{-1}$, corresponding to a X$_{\rm CO}$ factor of $4.1^{+5.9}_{-2.4}\times10^{20}$\,cm$^{-2}$\,(K\,\kms)$^{-1}$. The shaded area indicate the standard error of the fitting line.
The vertical line indicates the detection limit (more than five $\sigma$ over two velocity channels). The typical error is about a factor of two.
\label{xfactor}}
\end{figure}

Figure\,\ref{xfactor} shows the plot of CO luminosity and virial masses for the identified clouds in NGC\,5253.
We find a linear correlation between the virial masses and luminosities.
The intercept of the best fit in the log--log space is $9.2^{+13.2}_{-5.4}$, where the slope has been constrained to be unity, which yields a conversion factor of $4.1^{+5.9}_{-2.4}\times10^{20}$\,cm$^{-2}$\,{(K\,\kms)$^{-1}$}.
The conversion factor is comparable within the error bar, although probably larger than the typical Galactic standard value \citep[$X_{\rm CO}=2$--3;][]{1987ApJ...319..730S,2009ApJ...699.1092H,2013ARA&A..51..207B}, and similar to the median value in LMC (metallicity similar to that of NGC\,5253) at 11 pc scale, $X_{\rm CO}=4.7$ \citep{2010MNRAS.406.2065H}.
In Figure\,\ref{xfactor} we also highlight the clouds from N5253-D around the SB for comparison as we did in Figure\,\ref{larson} for the line width--radius correlation scaling law, but no particular difference was found this time.

We compare the derived $X_{\rm CO}$ factor with the well-known trend between the $X_{\rm CO}$ factor and metallicity \citep{1996PASJ...48..275A,2008ApJ...686..948B,2011ApJ...737...12L,2012AJ....143..138S,2016A&A...588A..23A,2010ApJ...716.1191W}.
The correlation between $X_{\rm CO}$ and metallicity is determined based on several methods: a comparison between CO intensity and virial mass \citep[virial method;][]{1996PASJ...48..275A}, dust-based modeling using IR observations \citep[][]{1997A&A...328..471I,2011ApJ...737...12L},  and an empirical relation between the H$_2$ depletion time, metallicity and specific SFR \citep[SFR scaling method;][]{2012AJ....143..138S,2016A&A...588A..23A}.
The trend of the CO-to-H$_2$ conversion factor as a function of metallicity is given by $\log_{10}X_{\rm CO}=A+N\times(12+\log_{10}{\rm O/H})$, where $A$ and $N$ are the intercept and slope of the correlation. 
Here, $A=9.30$ and $N=-1.0$ as determined from the virial method \citep{1996PASJ...48..275A}, while $A=13.30$, $N=-1.45$ from the SFR scaling method \citep[][]{2016A&A...588A..23A}.
Note, however, that the SFR scaling method used by \citet{2012AJ....143..138S} provides an $X_{\rm CO}$ factor several times larger at the lower metallicity end than that in \citet{2016A&A...588A..23A} due to a different set of assumptions.
As for the dust modeling method, since \citet{2011ApJ...737...12L} does not explicitly mention the equation to derive $X_{\rm CO}$ due to their low number of samples,  we derived $A=11.9$ and $N=-1.3$ by a linear fitting estimation.
It should also be noted that the dust modelling method used by \cite{1997A&A...328..471I}  again provides values a few times larger than those in \citet{2011ApJ...737...12L}.

The $X_{\rm CO}$ factor derived from the virial method accounts correctly for the mass of the CO-bright cores but may underestimate the mass of the whole H$_2$ envelopes \citep{2008ApJ...686..948B}.
 \citet{2007ApJ...656..168L} found that the metallicity of NGC\,5253 in the nuclear SB region was characterized by $12+\log({\rm O/H})\sim 8.18 \pm 0.03$, while $2\arcsec$--6$\arcsec$ southward from it by $12+\log({\rm O/H})\sim8.30\pm0.04$. Other metallicity measurements are consistent with these results \citep{1997ApJ...477..679K,1999ApJ...514..544K}.
With the equation above and the observed metallicities, we obtain $X_{\rm CO}$ factors for NGC\,5253 of (10--13), (14--21), and (18--27) $\times10^{20}$\,cm$^{-2}$\,{(K\,\kms)$^{-1}$}, using the virial, dust modeling, and SFR scaling methods, respectively.
Not surprisingly, our result is closer to the virial-method-based $X_{\rm CO}$ factor, but it is important to note that in general all these values are larger than our result. 
This is in agreement with a significant amount of gas not being accounted for by using the virial method $X_{\rm CO}$ factor, especially in the case of the low-metallicity environment of NGC\,5253.

Next we instead estimate the molecular mass of the clouds from the dust mass and obtain the $X_{\rm CO}$ factor.
\citet{2013A&A...557A..95R} estimated a dust mass $M_{\rm dust}=1.2^{+1.1}_{-0.6}\times10^5$\,$M_{\sun}$ (corrected by our assumed distance) within $120\arcsec$, equivalent to the whole galaxy by fitting the complete infrared spectral energy distribution.
Using the correlation between gas-to-dust mass ratio (GTD) and metallicity in \citet{2014A&A...563A..31R}, the GTD ranges from 398 to 966 for the metallicity of NGC\,5253, $12+\log({\rm O/H})\sim 8.18-8.30$.
This GTD yields a total gas mass of $M_{\rm gas}=M_{\rm dust}\times {\rm GTD} \simeq (5-12)\times10^7$\,$M_{\sun}$. 
The total gas mass can also be expressed as $M_{\rm gas}=\mu_{\rm gal}(M_{\rm H{\sc I}}+M_{\rm H_2})$, where $\mu_{\rm gal}$ is the mean atomic weight, which is 1.38 \citep{2014A&A...563A..31R}, $M_{\rm H{\sc I}}$ is the atomic gas mass, and $M_{\rm H_2}$ is the molecular gas mass.
An $M_{\rm H{\sc I}}$ of $1.03\times10^8$\,\Msol (also corrected by our assumed distance) was obtained with single-dish observations within $11\times7$\,arcmin$^2$  \citep{2012MNRAS.419.1051L}.
Assuming a Gaussian distribution of the \ion{H}{1} gas over the 77\,arcmin$^2$ area, the $M_{\rm H{\sc I}}$ within $120\arcsec$ is $1.4\times10^7$\,\Msol.
The total flux density of NGC\,5253 is $S_{\rm CO(2-1)}=92.6$\,Jy\,\kms\, within a radius of 22$\arcsec$, and we assume that it is not more extended than that.
The $X_{\rm CO}$ factor can then be obtained from equation $\mu_{\rm gal}M_{\rm H_2}=5.5\times10^{3}\,X_{\rm CO} D^2_{\rm L} S_{\rm CO(2-1)}{R_{21/10}}^{-1}$,
and with the assumptions above it yields a range of $X_{\rm CO} = 6-19$. 
This is about 1.5 to five times larger than the average we obtained, but it is still within the range of the estimated $X_{\rm CO}$ factors obtained using the dust model and SFR scaling methods. This supports that the net $X_{\rm CO}$ factor is likely larger if recovering the CO-dark H$_2$ gas.

\subsubsection{Surface Density and Gas Pressure}
\label{density}

\begin{figure}
\includegraphics[width=0.5\textwidth]{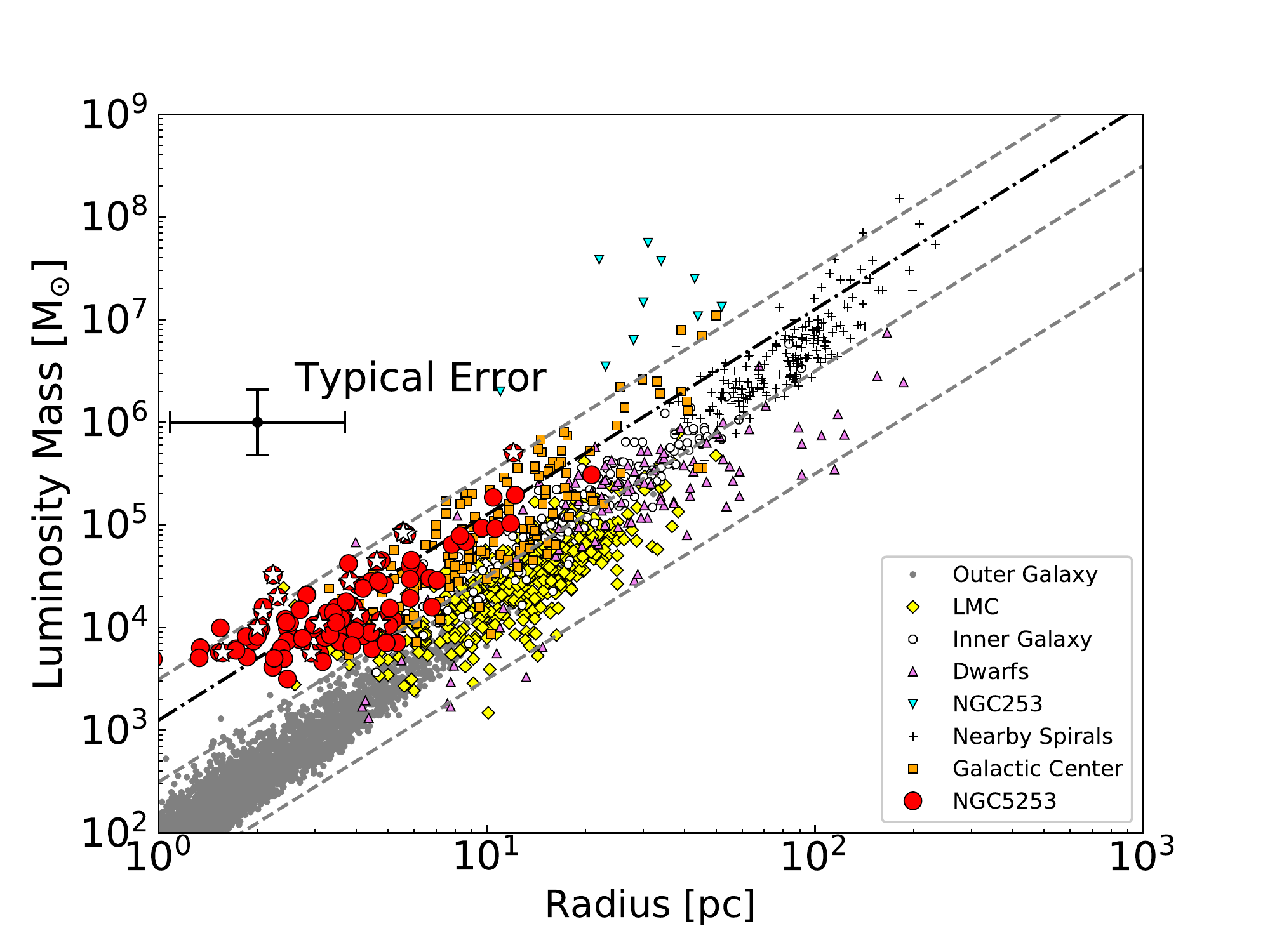}
\caption{Mass as a function of radius. The symbols are the same as Figure\,\ref{larson}. The dashed lines indicate 10, 100, and 1000\,$M_{\odot}\,$pc$^{-2}$, and the dashed-dot line 400\,$M_{\odot}\,$pc$^{-2}$. \label{massR}}
\end{figure}

Figure\,\ref{massR} shows the {\rm gas mass} as a function of radius for the molecular clouds of NGC\,5253 and compared with other nearby galaxies and our Galaxy.
The {\rm gas mass} was derived assuming $X_{\rm CO}=4.1$.
Although the scatter is relatively large, the majority of the NGC\,5253 clouds are aligned along the line of a surface density of $\Sigma_{\rm  H_2}$ = 400\,\Msol\,pc$^{-2}$, higher than the general trend for the molecular clouds in our Galaxy and other nearby galaxies.
This would be even higher if we had used a larger  $X_{\rm CO}$ factor.
The molecular clouds in NGC\,5253 follow the scaling relation of molecular clouds in the Galactic Center, especially those close to the central SB,  where data points are close to the line of surface density equal to $\Sigma_{\rm  H_2}$ = $10^3$\,\Msol\,pc$^{-2}$.

\begin{figure}
\includegraphics[width=0.5\textwidth]{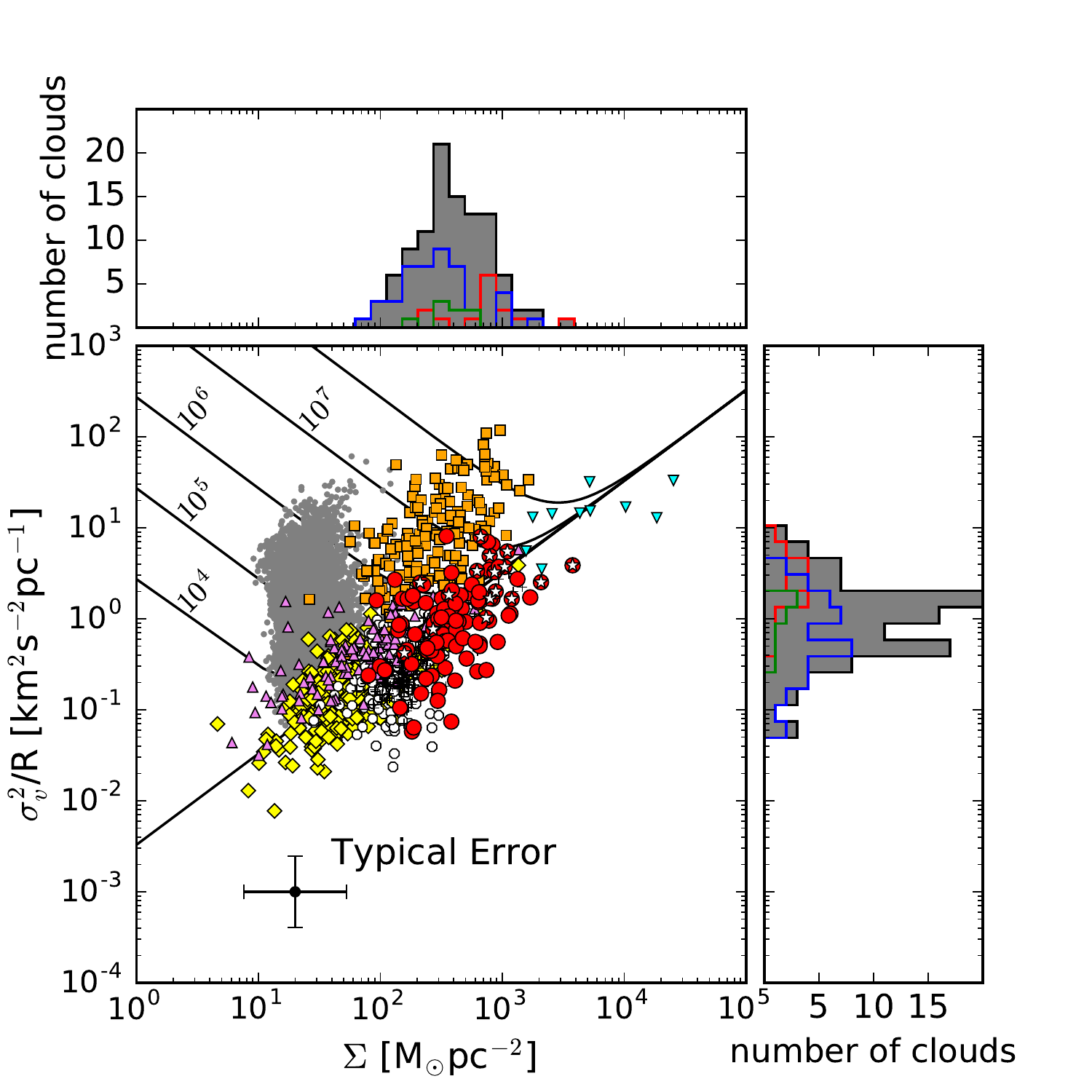}
\caption{
Scaling coefficient $\sigma_v^2/R$  as a function of surface mass density $\Sigma_{\rm  H_2}$. The symbols are the same as in Figure\,\ref{larson}. 
The curves show the equilibrium for external pressures with $P/k_B=10^4, 10^5,10^6, 10^7$, and $10^8$\,cm$^{-3}$\,K \citep{2011MNRAS.416..710F}.
The straight solid line indicates no external pressure, i.e. $P/k_B=0$\,cm$^{-3}$\,K.
The top and right panels are histograms of the scaling coefficient and surface density of the NGC\,5253 clouds, respectively. 
The gray-filled histogram represents all identified clouds and the ones in red, blue, and green lines are those in N5253-D, N5253-C, and N5253-F, respectively.  
\label{pressure}}
\end{figure}

Figure\,\ref{pressure} shows the correlation between the scaling coefficient $\sigma_v^2/R$ and the gas mass surface density,
calculated following the convention in \citet{2011MNRAS.416..710F}, in order to study the role that external pressure plays in confining molecular clouds.
The curves show the solutions for six different external pressures. The  straight line corresponds to no external pressure or a gravitationally bound state.
For example, the molecular clouds in the outer Galaxy and in the Galactic Center are located above the straight line, which indicates that 
these clouds are pressure-bound clouds.
The clouds in NGC\,5253 are clustered around the straight line as in other extragalactic molecular cloud complexes.
According to \citet{2011MNRAS.416..710F}, this suggests that external pressure is not needed to support the molecular clouds, and they would be in a gravitationally bound state.

The histograms in Figure\,\ref{pressure} show that the clouds close to the central SB (N5253-D) are characterized by relatively higher surface densities ($\Sigma_{\rm  H_2}$ $\sim10^3\,\Msol$\,pc$^{-2}$) and $\sigma_v^2/R$ ($\sim3$\,km$^2$\,s$^{-2}$\,pc$^{-1}$),
compared to other clouds in the same galaxy, although all clouds in the galaxy are likely in equilibrium.
In other words, the clouds near the central SB require higher internal pressure than the others to be gravitationally bound. 
The thermal pressure is roughly calculated to be $P/k_B\gtrsim10^6-10^7$ and $P/k_B\sim10^5-10^6$ for N5253-D and N5253-C, respectively, using the molecular gas density and kinetic temperature for N5253-D and N5253-C in \citet{2015Natur.519..331T}, that is, $T_k>200$\,K and $n_{\rm H_2}\sim(4.5\pm0.5)\times10^4$\,cm$^{-3}$, and 
 $T_k\sim15-20$\,K and $n_{\rm H_2}\sim(3.5-4)\times10^4$\,cm$^{-3}$, respectively.
Note that we cannot calculate this for N5253-F because there is no measurement of density or temperature to date.
The turbulence pressure for each molecular cloud complex is estimated to be similar using the molecular gas density and the typical velocity dispersion.
Thus the expected internal pressure can be explained by thermal pressure and turbulence.

Although the molecular clouds in the central SB (N5253-D) and in other regions (N5253-C/F) are gravitationally bound, the internal and gravitational pressures of the clouds near the central SB are systemically higher, with the median values of the central clouds shifted 0.5\,dex relative to the median values of the other clouds.

\section{Discussion}
\label{discussion}
\subsection{Association with SSCs and Feedback}

In this section, we discuss the star formation activities and the properties of clouds that appear associated with young clusters.
To do this, we compare the spatial association of the identified CO clouds with previously identified star clusters \citep{2015ApJ...811...75C,2013MNRAS.431.2917D,2004ApJ...603..503H}.
\citet{2015ApJ...811...75C} have revised age and mass estimations in \citet{1997AJ....114.1834C} toward 11 star clusters using multiple-wavelength data and a more accurate spectral energy distribution model fitting. 
We also used the catalog in \citet{2013MNRAS.431.2917D} to cover other star clusters. This includes 181 identified clusters in NGC\,5253 using UV/optical/NIR {\it HST} data, covering an area of about $500\times430$\,pc$^2$.
For sources that do not have any measurement in \citet{2013MNRAS.431.2917D} nor \citet{2015ApJ...811...75C}, we use the star cluster catalog by \citet{2004ApJ...603..503H}, obtained from optical broadband {\it HST} data.
The uncertainty of absolute astrometry of the {\it HST} image is $0\farcs1-0\farcs3$ and the two bright clusters in the {\it HST} image have a $0\farcs2$ offset from the two 1.3 cm radio components in \citet{2000ApJ...532L.109T} \citep{2015ApJ...811...75C}.
It is likely that the two bright clusters correspond to the two radio components and we correct the coordinates of all star clusters in the catalog accordingly.

\begin{figure*}
\includegraphics[width=.45\textwidth, trim={50 30 150 40}, clip]{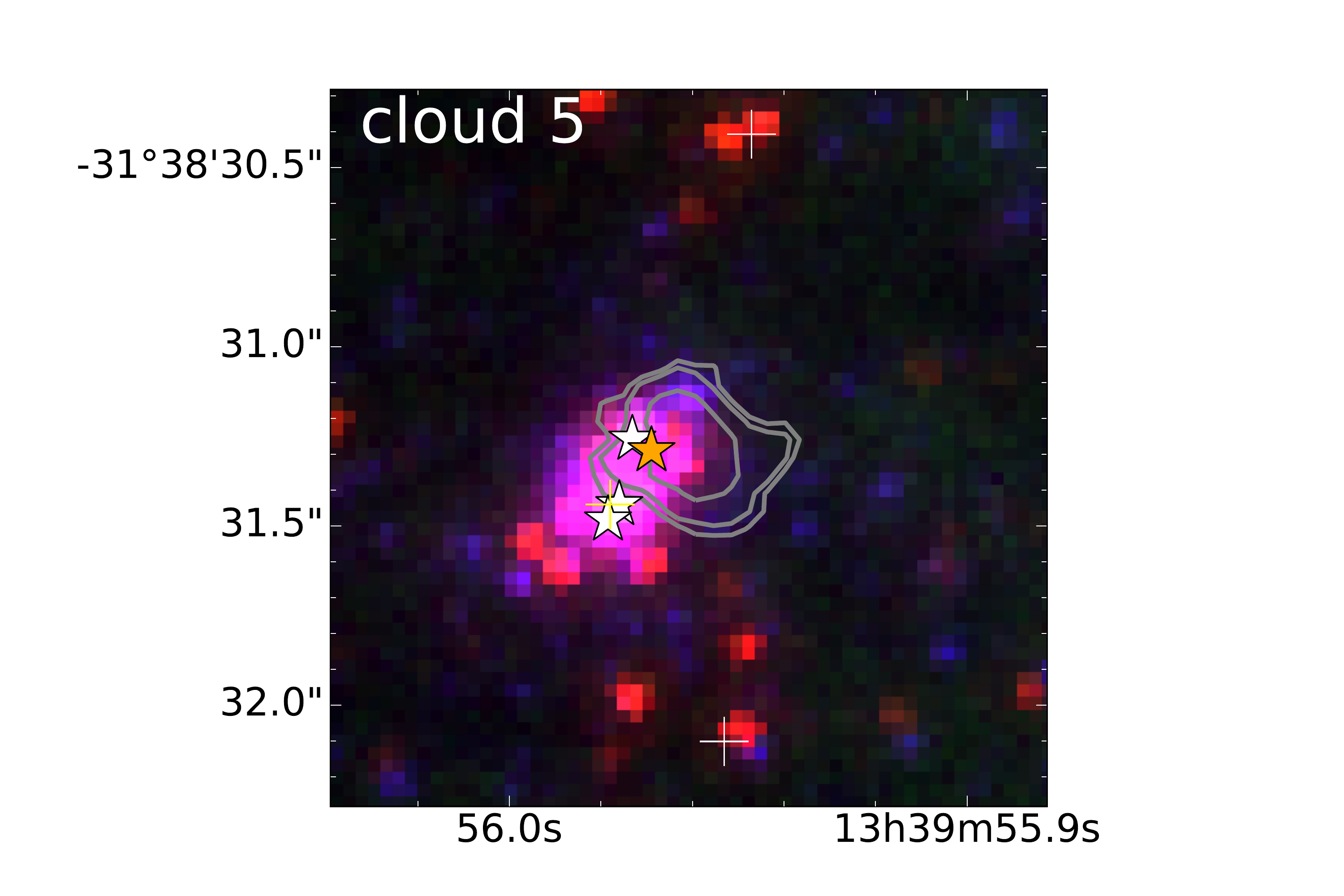}
\includegraphics[width=.45\textwidth, trim={50 30 150 40}, clip]{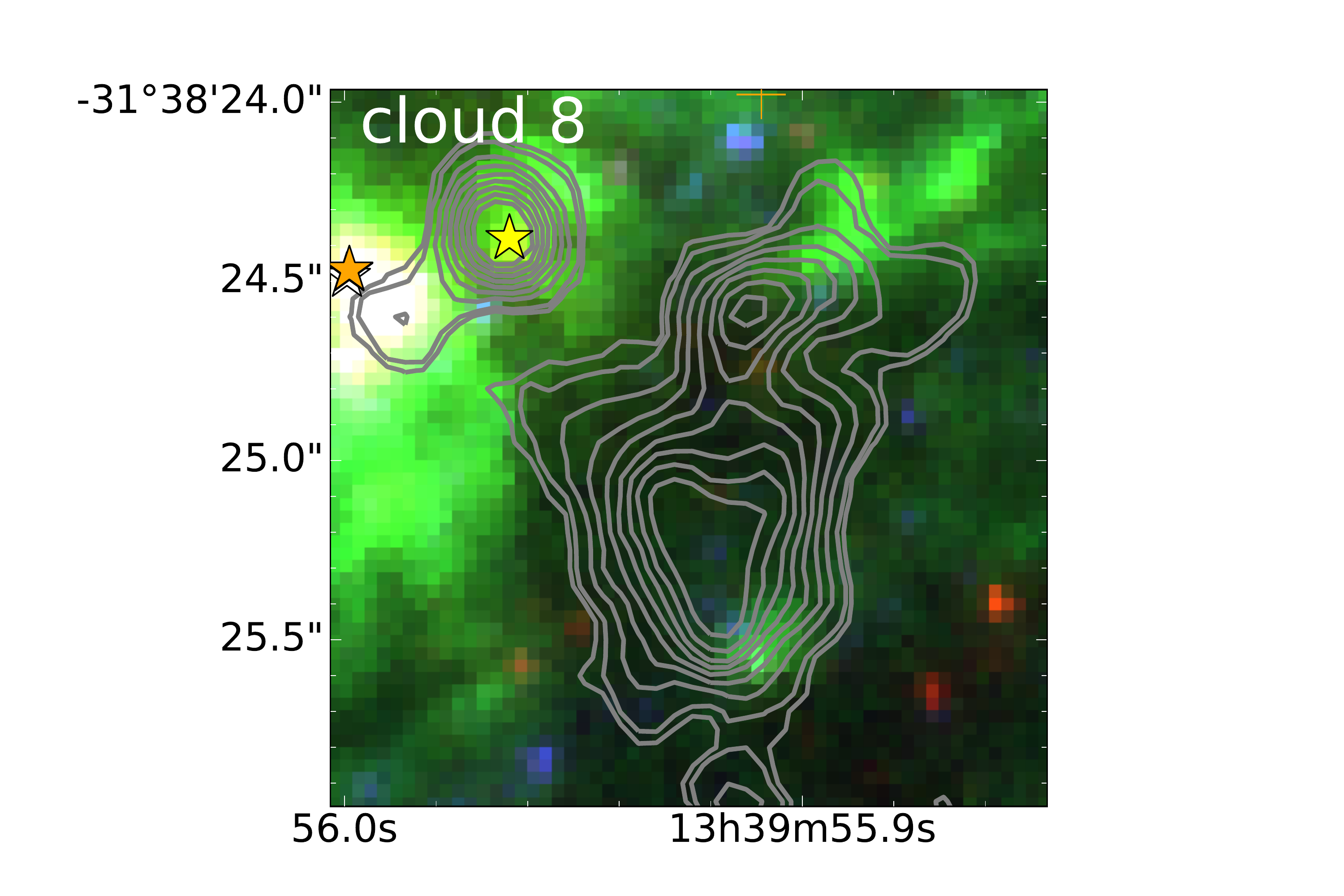}\\
\includegraphics[width=.45\textwidth, trim={50 30 150 40}, clip]{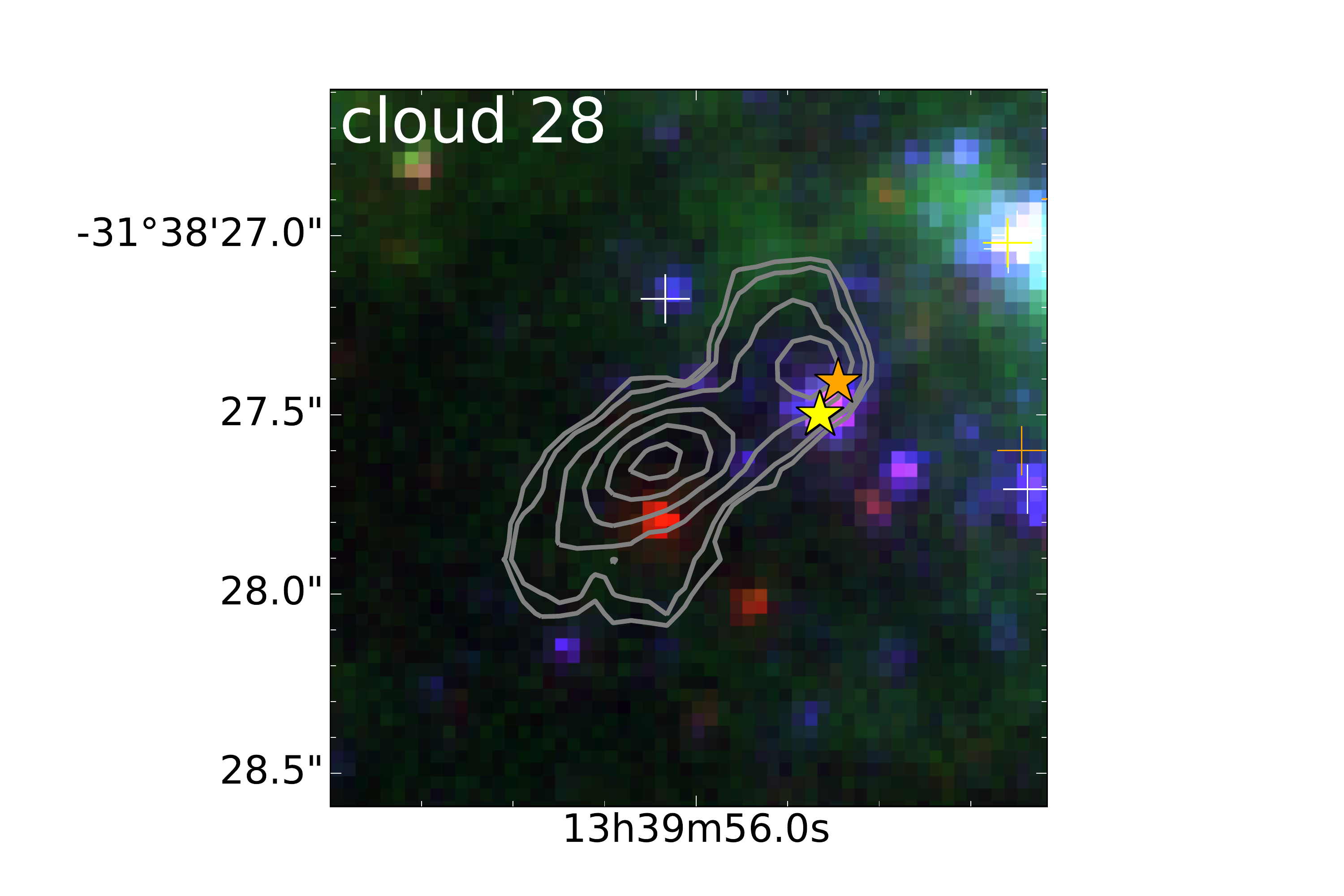}
\includegraphics[width=.45\textwidth, trim={50  30 150 40}, clip]{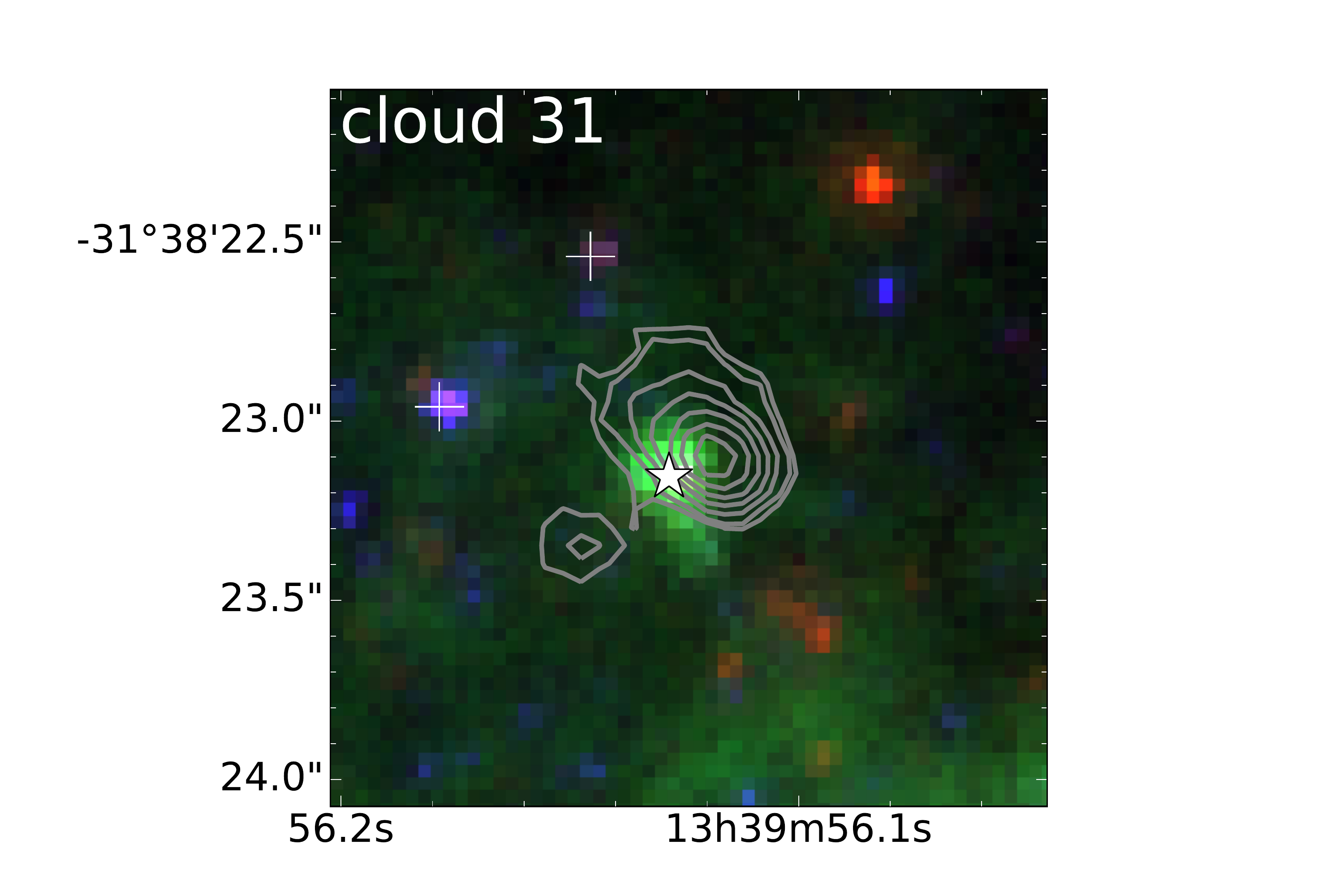}\\
\includegraphics[width=.45\textwidth, trim={50 15 150 40}, clip]{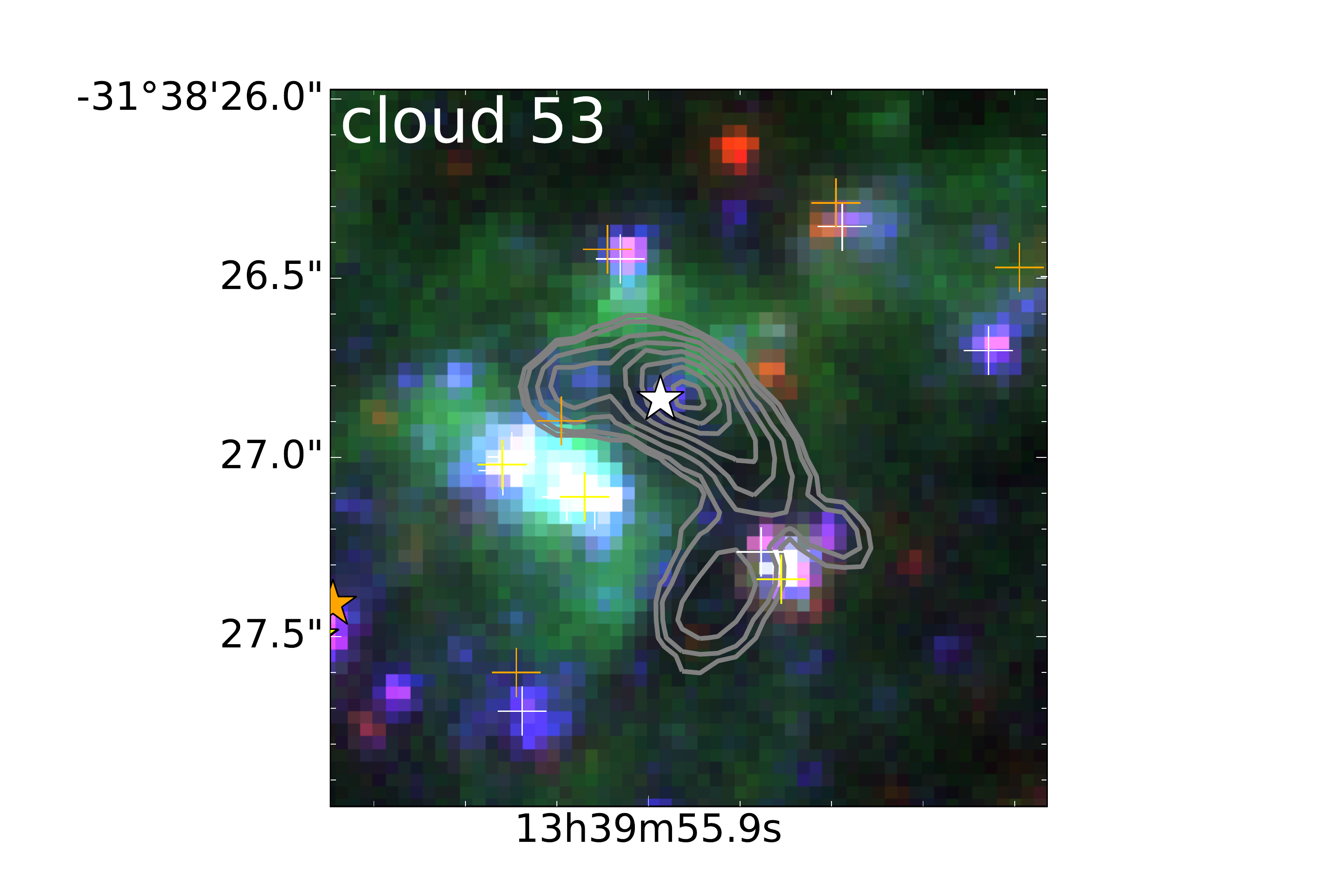}
\includegraphics[width=.45\textwidth, trim={50 15 150 40}, clip]{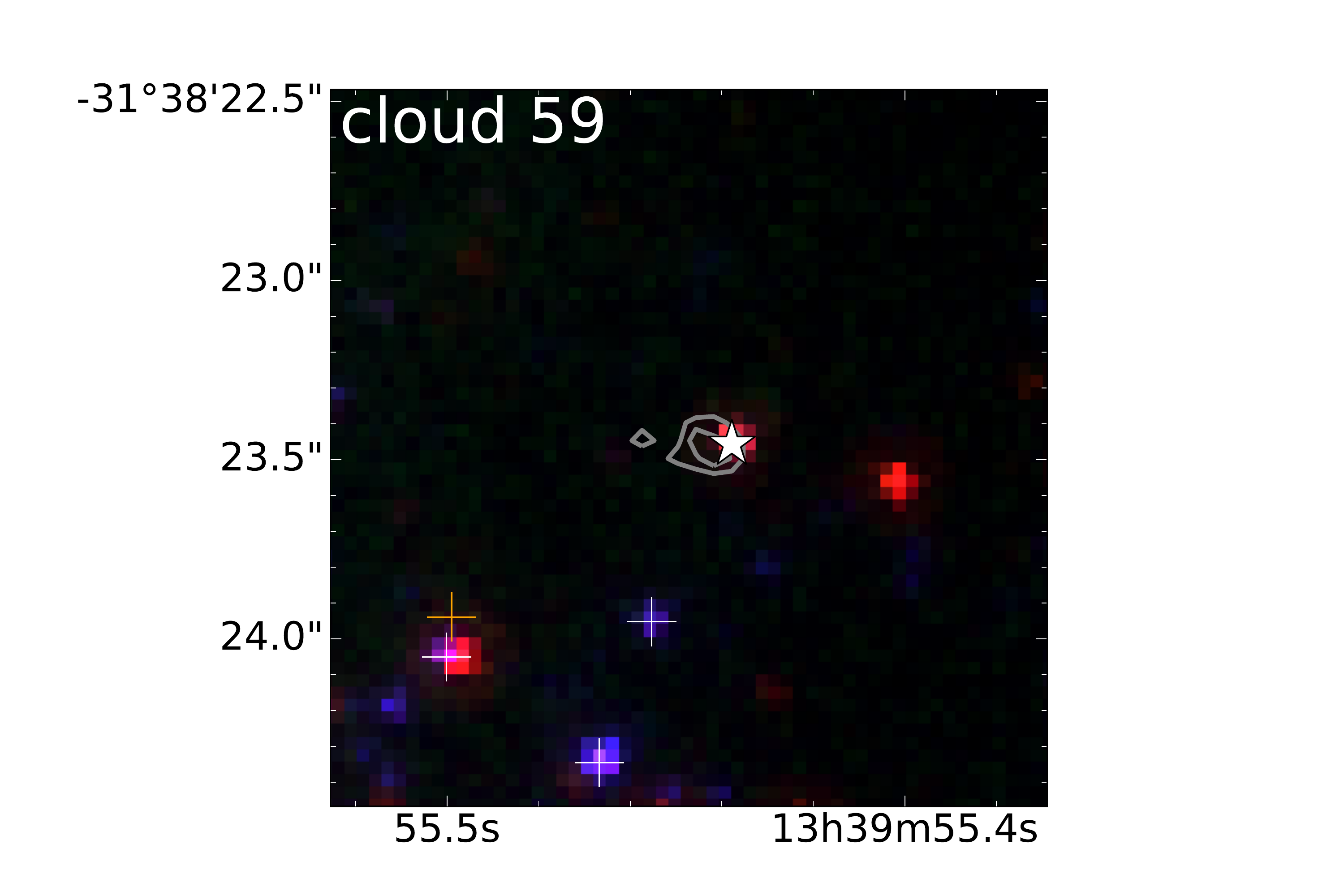}
\caption{{\footnotesize CO integrated intensity map (contours) for individual clouds\,5, 8, 28, 31, 53 and 59 over the {\it HST} composite images. The contour levels are 3, 5, 10, 15, 20, 25, 30, 35 and 40\,$\sigma$, where $\sigma$=0.01\,Jy\,\kms. The RGB {\it HST} composite images are composed of F300W in blue, F658N or H$\alpha$ in green, and F814W in red. The star symbols indicate the star clusters which are associated with the molecular clouds and each color (white, yellow, orange) indicates the different star cluster catalog (\citealt{2013MNRAS.431.2917D}, \citealt{2015ApJ...811...75C}, \citealt{2004ApJ...603..503H}, respectively). The cross symbols are star clusters in our compilation which are not associated with any of the identified clouds. All panels are $2\arcsec$ $\times$ $2\arcsec$.
}\label{starcloud1}}
\end{figure*}

After correcting the coordinates of the clusters, we compare their location with that of our identified clouds. 
The {\it HST} images also cover most of the observed field in CO (N5253-D, N5253-F and the western half of N5253-C).
We search for clusters whose central position spatially overlaps with the extent (at the level of $3\,\sigma$) of any of the molecular clouds, taking into account the uncertainty of astrometry.
Among the 118 clouds, we found that six clouds appear spatially associated with 10 star clusters.
Figure\,\ref{starcloud1} shows the CO integrated intensity images of the six clouds overlaid on their corresponding {\it HST} composite image.
The identifications of these clouds, their associated star clusters, and their properties are listed in Table \ref{sfe}.
The majority of the associated clusters are young, less than 10\,Myr, which is the maximum age acceptable to infer that they may have originated from the clouds, taking into account the lifetime of a giant molecular cloud \citep{2009ApJS..184....1K,2012ApJ...761...37M,2011ApJ...729..133M}.
Specifically, in the case of cloud\,8, we found that its velocity and location is very close to those of the H30$\alpha$ peak or the central cluster  \citep{2017MNRAS.472.1239B,2000ApJ...532L.109T,2004ApJ...610..201S}, which together with the young age of the cluster likely suggest a physical relation between the cloud and the star clusters.
Note that cluster G-9, which is paired with cloud\,60, has a large uncertainty in age, which can be up to 100\,Myr depending on the stellar synthesized model that is used.
If G-9 is that old, it probably did not originate from that molecular cloud.

We estimated the SFEs for these clouds.
SFE is defined as $M_{\ast}/(M_{\rm gas}+M_{\ast})$, where $M_{\ast}$ is the mass of a star cluster, and $M_{\rm gas}$ is the mass of the cloud.
The derived SFE is also listed in Table \ref{sfe}. The SFEs we obtain for NGC\,5253 range from 5\,\% to 64\,\%.
Compared with cluster-forming clouds in our Galaxy, in which the SFE ranges from 10\.\% to 30\,\% \citep{2003ARA&A..41...57L},
two among all, cloud\,5 and cloud\,8, show high SFE. 
Note that we exclude the case of cloud\,60 here because of the large uncertainty in its age, as explained above.
Cloud\,8 has clumpy structure, and if one of its clumps is associated with one of the two clusters (cluster\,11), the SFE would be 80\,\%.

In the SSC-forming clouds in an SB system, whether gas collapses or disperses is a competitive process between the inward forces such as gravity and outward forces such as that caused by a jet, expansion of a \ion{H}{2} region, stellar wind, and radiation pressure.
The most dominant force in the SSC-forming clouds is radiation force ($F_{\rm rad}$) among the outward forces and gravity ($F_{\rm grav}$) for the inward forces  \citep{2010ApJ...709..191M}.
Following \citet{2011ApJ...729..133M}, we estimate $F_{\rm rad}$ for five clouds using 
$F_{\rm rad}=L/c$\,dynes,
where $L$ is the luminosity in erg\,s$^{-1}$, and $c$ is the light speed in cm\,s$^{-1}$.
We use $L=\xi Q$, where $Q$ is the ionizing photon rate and $\xi=8\times10^{-11}$\,ergs \citep{2010ApJ...709..191M}, and the conversion  between $M_{\ast}$ and $Q$ is
$M_{\ast}=1.6\times10^4 (Q/10^{51}\,{\rm s}^{-1})\,\Msol$ \citep{2011ApJ...729..133M}.
In this conversion between $M_{\ast}$ and $Q$,  \citet{2011ApJ...729..133M} assumed a minimum stellar mass of $0.1\,M_{\odot}$, a maximum stellar mass of $120\,M_{\odot}$, a slope of the initial mass function of $-1.35$ and constrained to young star clusters ($\sim4$\,Myr, the averaged main-sequence lifetime of massive stars, which are mainly responsible for emitting ionizing photons). 
 For the force of gravity, we use $F_{\rm grav}=GM_{\rm gas}^2R^{-2}$, where $R$ is the cloud radius \citep{2011ApJ...729..133M}.

Table\,\ref{sfe} shows the estimated forces for each cloud and star cluster.
Since the conversion from $M_{\ast}$ to $Q$ is limited to young stellar age populations, the $Q$ might be overestimated in the case of an older stellar cluster \citep[beyond $\sim5$\,Myr;][]{2011ApJ...729..133M}, and thus for clouds\,5, 53 and 59, $F_{\rm rad}$ would be an upper limit. Note that we calculate $F_{\rm grav}$ both using $M_{\rm gas}$ only and $M_{\rm gas}+M_{\ast}$ (as shown in the parentheses in the table). 
In the latter, it is assumed that the star clusters are embedded inside the molecular cloud. 
In the case of cloud\,8, this scenario is plausible since its associated star clusters are still very young and located at one of the CO emission peaks. 
Besides, the virial mass of cloud\,8 is much larger when compared to the gas mass only, but it is comparable to the summation of the stellar clusters and the gas masses, unlike in other clouds.
Note that for cloud\,5 and cloud\,53, since the ages of the associated clusters G-125, G-127, and G-86 are not available, we used values from only clusters H-2 and C-4, respectively, and thus $M_{\ast}$ and consequently $F_{\rm rad}$ are a lower limit.
If only part of cloud\,8 is associated with the star cluster, the radiation force would be more than five times larger than the gravitational force.
Clouds\,5 and 8 (if only part of cloud\,8 is associated with the star cluster) are unique clouds in the sense that the radiation force exceeds the inward force of gravity, while the rest of the clouds are governed by gravity. This means that these two clouds will likely dissipate the parental clouds, due to stellar feedback after the SSC is formed,
while the other clouds (e.g. clouds\,28, 31, and 53) have still have the possibility to form stars in the gravity-dominant phase.
Cloud\,5 is associated with a cluster with an age of 10\,Myr, while cloud\,8 is associated with a younger cluster (1\,Myr), which is the most massive cluster in the galaxy.
Taking into account that the clouds in the central SB region have high gas densities, it is reasonable to expect that 
a more massive cluster will tend to be formed, but then its destructive stellar feedback will remove the gas quickly.
Note that for an $X_{\rm CO}$ factor 2.6 times larger than the value applied here, all clouds would be in the gravity-dominant phase. In this case, the parental molecular clouds would have enough content to form more stars rather than being dissipated.

\subsection{Origin of the Starburst: Comparison of the Global Gas Distribution with Simulations}
NGC\,5253 is argued to have experienced external gas infall because the kinematics of the atomic and molecular gas cannot be explained by galaxy rotation.
The gas infall has likely triggered the powerful SB \citep{2002AJ....124..877M,2015PASJ...67L...1M,2012MNRAS.419.1051L,2014A&A...566A..71L}.

\citet{2014MNRAS.442.1830V} showed that external gas infall can trigger a starburst in dwarf galaxies.
 Those simulations show that the SBs can happen several times within 1\,Gyr depending on the initial conditions.
 Once external gas encounters the dwarf galaxy, the kinematics/distribution of the gaseous component in the host galaxy is strongly disturbed. The gas loses momentum and is driven inward to form dense clouds and consequently cause the SB.  
 The new stars blow out the surrounding gas to large radii, but it falls back again into the galaxy to be the fuel to form the next SB event. 
  The simulations also predict that the gaseous and stellar components of the galaxy become more compact and result in higher densities toward the center.
As the burst settles down, the rotation curve becomes steep, as observed in several BCDs \citep{2001AJ....122..121V,2014MNRAS.441..452K}.

From the CO observations, we do not find any clear sign of simple disk rotation, in agreement with previous \ion{H}{1} observations \citep{2012MNRAS.419.1051L}. The distribution of CO emission is very irregular, and the direction of the molecular gas structure does not match that of the H$\alpha$ outflow. This support the scenario that the gas might be mostly infalling to the center and the SB event is most likely recent.  The central SB region has high surface gas densities (Section\,\ref{density}), which supports simulation results that show that before and during the burst, gas flows to the center and the density rises significantly compared to larger radii. 
Taking into account that several bubbles caused by expansion of \ion{H}{2} regions and supernovae are found within the galaxy \citep[][]{1998AJ....116.1212T,1997AJ....114.1834C,1999MNRAS.306...43S} and \ion{H}{1} emission follows (although the resolution is much coarser) the H$\alpha$ bubbles \citep{2012MNRAS.419.1051L}, part of these molecular gas components may also be gas that is blown out to large radii and falling back to the galaxy.

First, we calculate the gas inflow rate through the dust lane (i.e. N5253-C). 
If the gas of N5253-C were infalling, as suggested in previous studies, and using a mass of $M_{\rm gas}$ = $7.6\times10^6\,\Msol$ (equivalent to the gas mass derived from the emission within the ellipse of N5235-C in Figure\,\ref{mom}), the velocity difference of 30\,\kms\ between N5253-C and the central SB as the radial inflow speed, and the distance of N5253-C to the central SB, $\sim$100\,pc, we obtained a gas inflow rate via the dust lane of $\dot{M}_{\rm flow}$ $\simeq$2\,\Msol yr$^{-1}$. 
 This was obtained by using equation $\dot{M}_{\rm flow}$[\Msol\,yr$^{-1}$]= $1.0\times10^{-6} M_{\rm gas} V_{\rm grad, corr}$.
Here, $V_{\rm grad,corr}$ is the velocity gradient in units of $\kms\,$pc$^{-1}$, and given by $ V_{\rm grad,corr}  =V_{\rm grad}/\tan i$, where $i$ is an inclination of the flow along the line of sight, and we assume  $i=45$\arcdeg.
A caveat with the infalling scenario is that in the data presented here, we have not found molecular gas features in the PVDs clearly connecting the dust lane (N5253-C) and the central SB (N5253-D).

We show in Figure\,\ref{pvcut} the expected P--V curves for a disk-like model assuming a maximum rotation velocity of 30\,\kms\ (Appendix\,\ref{a1}). Although the kinematics of some of the structures may be explained by galaxy rotation, in general this is not true, and the difference is most noticeable for components at PAs 70$^{\arcdeg}$, 100$^{\arcdeg}$ and 175$^{\arcdeg}$. We have found from the PVDs that multiple components connect to the central massive molecular cloud (cloud\,8), showing a smooth velocity gradient (Section\,\ref{filament}).

We measure the total gas mass along these structures, $M_{\rm gas}$, and the velocity gradient, $V_{\rm grad}$, in order to estimate the gas flow rate as done before for the dust lane.
The obtained gas flow rate is about 0.06--0.07\,\Msol yr$^{-1}$ each (Table\,\ref{outflow}).
If these are all inflowing, then the accumulated gas would be $1.9\times10^5$\,\Msol\, in 1\,Myr. 
To form a molecular cloud like cloud\,8 only from these gas inflows, it would take $\sim10$\,Myr, which is an equivalent timescale for the formation of a molecular cloud of $<20$\,Myr \citep{1994LNP...439...13L}.
Note that the calculated inflow rate is lower than the previously estimated value in \citet{2015PASJ...67L...1M}, because it was assumed that the whole molecular cloud complex N5253-D contributed to the inflow.

The SFR of NGC\,5253 in the literature is in the range of 0.1--0.4\,\Msol yr$^{-1}$ \citep{2000ApJ...532L.109T,2002AJ....124..877M,2004AJ....127.1405C,2012MNRAS.419.1051L,2015ApJ...811...75C,2017MNRAS.472.1239B}.
Even taking into account that other possible mass losses such as mass ejection in the form of hot gas \citep[$\sim$0.2\,\Msol yr$^{-1}$;][]{2004MNRAS.351....1S} and ionized gas outflow \citep[(0.4--2.6)\,$\times 10^{-3}$\,\Msol yr$^{-1}$;][]{2015PASJ...67L...1M}, the gas inflow rate is much larger than the total amount of outflow rate and SFR, which indicates that enough molecular gas is being fueled to maintain star formation. That there is enough amount of gas being fueled to the center may support \citet{2015ApJ...811...75C}, who suggested the continuous star formation at least over 15\,Myr in the region within the central 300\,pc of NGC\,5253.

The multiple components connecting to the central massive molecular cloud, which show a smooth velocity gradient and deviate from galaxy rotation, might be due to gas that has been blown out once and rain down on the galaxy. Even if these small components were all outflows rather than inflows, a positive net inflow rate is likely preserved, because 
in this case the molecular outflow rate would only be 0.2\,\Msol yr$^{-1}$ (i.e. the summation of the flow rates via all three identified structures in the PVDs).

\section{Summary}
We presented the highest resolution and sensitivity ALMA CO(2--1) data to date toward the nearest young SB dwarf galaxy, NGC\,5253. We combined ALMA 12m, 7m and TP array data in order to recover all spatial scales. We summarize our main results as follows:

\begin{enumerate}
\item We have successfully resolved using ALMA observations the previously identified molecular cloud complexes into parsec-scale molecular clouds and also revealed the diffuse molecular emission scattered over the entire galaxy. 

\item We have identified a total of 118 molecular clouds within these molecular cloud complexes with typical sizes of 4.3\,pc and velocity dispersions of 2.2\,\kms. This is similar to the properties of molecular clouds in other galaxies and in our Galaxy. It is remarkable that the properties of the molecular clouds close to the central SB are found to be different than the rest; that is, they are characterized by large velocity widths and high gas surface densities, which may show the necessary conditions of the molecular gas to be able to form SSCs.

\item We have derived for the first time in this galaxy the $X_{\rm CO}$ factor, $X_{\rm CO}$ = $4.1^{+5.9}_{-2.4}\times10^{20}$\,cm$^{-2}$(K\,\kms)$^{-1}$ based on the assumption that the molecular clouds found in NGC\,5253 are virialized. This conversion factor would be a lower limit if the CO-dark H$_2$ gas is taken into account.

\item We found that the majority of clouds in NGC\,5253 are offset on average by 0.2\,dex from the general line-width - size relation found in our Galaxy and other nearby galaxies including dwarf galaxies, although not as much as in the Galactic Center environment. Molecular clouds close to the SB of NGC\,5253 present systematically the largest offsets, 0.5\,dex.

\item NGC 5253 clouds are aligned along the line of a surface density of $\Sigma_{\rm H_2}$ $\sim$ 400\,M$_\odot$\,pc$^{-2}$, higher than the general trend for the molecular clouds in our Galaxy and other nearby galaxies. In general, external pressure is not needed to support the molecular clouds, and in principle they would be in a gravitationally bound state, although exceptions exist close to the SB region. The clouds close to the central SB (N5253-D) are  characterized by relatively higher surface densities ($\Sigma_{\rm H_2}$ $\sim$ 10$^3$\,M$_\odot$\,pc$^{-2}$) and $\sigma_v^2/R$ ($\sim$ 3 km$^2$\,s$^{-2}$ pc$^{-1}$). 

\item  Based on the spatial comparison between the molecular clouds and the optically identified star clusters, we identified six clouds associated with young star clusters. Compared with cluster-forming clouds in our Galaxy, in which the SFE ranges from 10\,\% to 30\%, two among all, clouds 5 and 8, show very high SFEs, reaching up to 80\% (if one of the clumpy structures of cloud\,8 is associated with cluster 11).
Such high SFEs could be overestimated because these two clouds formed relatively massive star clusters and might be in the gas-removal phase due to the stellar feedback from these clusters. The fate of the cloud highly depends on the strength of the stellar feedback.

\item  We showed that the kinematics of the molecular gas is not likely dominated by the galaxy rotation of NGC\,5253, and that an asymmetric distribution and noncircular motions are present. Based on its distribution and kinematics, we interpret the molecular gas to be falling toward the center.
Compared with numerical simulations \citep{2014MNRAS.442.1830V}, this is compatible with an early phase of gas falling toward the center, where the distribution is rather chaotic and high surface densities are formed preferentially toward the center, which then form the SB with SSCs. Several structures that show the velocity gradient are found in the PVDs, which we interpret as inflows of material raining down on the galaxy, probably as a result of the SB event. 
Even if they were outflows, the total inflow rate would likely be large enough to sustain the continuous star formation in NGC\,5253.
\end{enumerate}

\acknowledgments
This paper makes use of the following ALMA data:
ADS/JAO.ALMA\#2013.1.00210.S. ALMA is a partnership
of ESO (representing its member states), NSF
(USA), and NINS (Japan), together with NRC (Canada),
NSC and ASIAA (Taiwan), and KASI (Republic of
Korea), in cooperation with the Republic of Chile.
The Joint ALMA Observatory is operated by ESO,
AUI/NRAO and NAOJ.
R.E.M. was supported by the ALMA Japan Research Grant of NAOJ Chile Observatory, NAOJ-ALMA-0073.
Data analysis was in part carried out on the open use data analysis computer system at the Astronomy Data Center (ADC) of the National Astronomical Observatory of Japan.
This research made use of {\sf Astropy}, a community-developed core Python package for Astronomy \citep{2013A&A...558A..33A}.

{\it Facilities:} \facility{ALMA}.

\appendix
\section{Disk Rotation}\label{a1}
 We estimate how different the kinematics is from simple disk rotation.
First we made a disk model with a rotation curve that is characterized by a constant velocity beyond $1\arcsec$ from the center, and at inner radii decreasing linearly to zero at the center.
According to the optical image, NGC\,5253 has the form of a disk galaxy (see Figure\,\ref{almafov}). We obtained the properties of this optical disk by fitting a two-dimensional Gaussian function.
The obtained parameters are listed in Table\,\ref{opt}.
The optical center is located in the middle of the three molecular cloud complexes (see Figure\,\ref{mom2}).
The systemic velocity of the stars in the center is obtained from optical spectroscopic observations \citep{2004ApJ...610..201S} and we assume the gas in the center also shares this velocity. 
We then determine the best-fit rotation velocity to be about 30\,\kms\, in order to reproduce the velocity field of the molecular gas, after fixing other parameters listed in Table\,\ref{opt}.
The comparison of the observed PVDs and the model for different PAs is presented in Figure\,\ref{pvcut}.

\begin{deluxetable}{lc}
\tablecaption{Basic properties of the optical disk of NGC\,5253\label{opt}}
\tablewidth{0pt}
\tablehead{
\colhead{Name} & \colhead{NGC\,5253} }
\startdata
Position Angle & 46\arcdeg\\
Optical diameter & 547\arcsec$\times$129\arcsec\\
Inclination & 77\arcdeg\\
Systemic velocity (radio, LSRK) & 389\,\kms\\
Rotation velocity &30\,\kms
\enddata
\end{deluxetable}

\begin{deluxetable}{lcrcl}
\tabletypesize{\scriptsize}
\tablecaption{Summary of ALMA Data\label{tbl1}}
\tablewidth{0pt}
\tablehead{
\colhead{Execution block IDs} & \colhead{Observation date} & \colhead{Time on source} & \colhead{Configuration} & \colhead{Data reduction\tablenotemark{a}} 
}
\startdata
uid://A002/X8440e0/X29c6 & 2014 Jun 15 &30.7\,minutes&ACA/7m&Manual (CASA 4.7.0)\\
uid://A002/X9652ea/X5c3  & 2014 Dec 10 &30.7\,minutes&ACA/7m&Manual (CASA 4.7.0)\\
uid://A002/X966cea/X25de  &  2014 Dec 11 &8.5\,minutes& C34-2/1&Manual (CASA 4.7.0)\\
uid://A002/Xa5df2c/X50ce  &  2015 Jul 18 &13.2\,minutes&C34-7/6&Manual (CASA 4.7.0)\\
uid://A002/Xa5df2c/X52fa  &  2015 Jul 18 &15.9\,minutes&C34-7/6&Manual (CASA 4.7.0)\\
uid://A002/Xa2300a/X12d5& 2015 Jun 05 &18.9\,minutes& ACA/TP & Pipeline (Cycle4-R2 on CASA 4.7.0) \\
uid://A002/Xa2300a/X95& 2015 Jun 03 & 18.9\,minutes&ACA/TP & Pipeline (Cycle4-R2 on CASA 4.7.0) \\
uid://A002/Xa2300a/X884& 2015 Jun 04 &18.9\,minutes& ACA/TP & Pipeline (Cycle4-R2 on CASA 4.7.0) \\
uid://A002/Xa24618/X1e5& 2015 Jun 04 & 15.4\,minutes&ACA/TP & Pipeline (Cycle4-R2 on CASA 4.7.0) \\
uid://A002/Xa24618/X53a& 2015 Jun 04 & 18.9\,minutes&ACA/TP & Pipeline (Cycle4-R2 on CASA 4.7.0) \\
uid://A002/Xa24618/X859& 2015 Jun 05 & 18.9\,minutes&ACA/TP & Pipeline (Cycle4-R2 on CASA 4.7.0) \\
uid://A002/Xa24618/Xb9d& 2015 Jun 05 &18.9\,minutes &ACA/TP & Pipeline (Cycle4-R2 on CASA 4.7.0) 
\enddata
\tablenotetext{a}{Data were calibrated manually or with the ALMA Pipeline. The CASA and ALMA Pipeline version used for calibration is indicated in parentheses. }
\end{deluxetable}

\clearpage
\LongTables
\begin{deluxetable*}{ccrcrrrrrrrrrr}
\tabletypesize{\scriptsize}
\tablecaption{Properties of Molecular Clouds in NGC\,5253 \label{tbl2}}
\tablewidth{0pt}
\tablehead{
\colhead{ID} & \colhead{($\Delta\alpha, \Delta\delta$)\tablenotemark{a}} & \colhead{$v_{\rm LSR}$} &
\colhead{$\sigma_v$}& \colhead{$\sigma_{\rm maj}\times \sigma_{\rm min}$ (P.A.)\tablenotemark{b}} &\colhead{$R$\tablenotemark{c}} & \colhead{$S_{\rm CO(2-1)}$} & \colhead{$M_{\rm vir}$\tablenotemark{c}} \\
\colhead{} & \colhead{($\arcsec$, $\arcsec$)} & \colhead{(\kms)} &
\colhead{(\kms)}& \colhead{(pc)}& \colhead{(pc)}  & \colhead{(Jy\,\kms)} & \colhead{(10$^4\,M_{\sun}$)} 
}
\startdata
1 & ($-4.2,  1.3$) & 356.& 4.0 $\pm$ 1.6 & $ 4.2\times 1.3$ ($18\arcdeg$) &\nodata& 0.21 $\pm$ 0.10 & \nodata \\ 
2 & ($-6.3, -1.3$) & 368.& 4.5 $\pm$ 0.3 & $ 9.1\times 3.6$ ($32\arcdeg$) & 10.5 $\pm$  0.7 & 3.32 $\pm$ 0.12 & 21.6 $\pm$  4.0 \\ 
3 & ($-3.6, -0.9$) & 368.& 2.7 $\pm$ 0.5 & $ 6.6\times 1.6$ ($-76\arcdeg$) &  4.6 $\pm$  0.7 & 0.77 $\pm$ 0.09 &  3.4 $\pm$  1.4 \\ 
4 & ($-6.9, -2.5$) & 366.& 0.9 $\pm$ 1.5 & $ 2.3\times 1.7$ ($-53\arcdeg$) &  2.7 $\pm$  2.7 & 0.14 $\pm$ 0.19 &  0.2 $\pm$  0.7 \\ 
5 & ($-1.2, -1.3$) & 368.& 2.3 $\pm$ 0.5 & $ 3.3\times 2.1$ ($77\arcdeg$) &  4.2 $\pm$  1.1 & 0.47 $\pm$ 0.10 &  2.4 $\pm$  1.4 \\ 
6 & ($-7.3, -3.6$) & 374.& 4.0 $\pm$ 0.8 & $17.1\times 2.0$ ($35\arcdeg$) &  9.7 $\pm$  1.3 & 1.67 $\pm$ 0.29 & 16.3 $\pm$  7.9 \\ 
7 & ($-7.0, -4.8$) & 373.& 3.7 $\pm$ 1.5 & $ 1.8\times 1.3$ ($-79\arcdeg$) &\nodata& 0.14 $\pm$ 0.09 & \nodata \\ 
8 & ($-1.7,  5.1$) & 398.& 8.1 $\pm$ 0.4 & $ 8.1\times 5.2$ ($23\arcdeg$) & 12.1 $\pm$  0.4 & 8.99 $\pm$ 0.16 & 83.1 $\pm$ 10.4 \\ 
9 & ($ 8.3, -0.2$) & 375.& 1.2 $\pm$ 1.9 & $ 4.1\times 2.0$ ($8\arcdeg$) &  4.5 $\pm$  3.3 & 0.11 $\pm$ 0.21 &  0.6 $\pm$  2.1 \\ 
10 & ($ 5.6,  7.1$) & 377.& 1.7 $\pm$ 1.0 & $ 1.7\times 1.6$ ($37\arcdeg$) &  1.7 $\pm$  2.5 & 0.11 $\pm$ 0.13 &  0.5 $\pm$  0.9 \\ 
11 & ($-4.8, -6.5$) & 379.& 4.3 $\pm$ 2.2 & $ 2.4\times 1.8$ ($-78\arcdeg$) &  2.8 $\pm$  2.2 & 0.37 $\pm$ 0.29 &  5.4 $\pm$  8.2 \\ 
12 & ($-5.4, -1.4$) & 376.& 1.4 $\pm$ 2.1 & $ 2.1\times 2.2$ ($44\arcdeg$) &  3.1 $\pm$  3.2 & 0.17 $\pm$ 0.26 &  0.6 $\pm$  1.9 \\ 
13 & ($-0.1, -8.5$) & 382.& 1.4 $\pm$ 0.6 & $ 2.8\times 2.1$ ($-47\arcdeg$) &  3.8 $\pm$  1.7 & 0.23 $\pm$ 0.11 &  0.8 $\pm$  0.8 \\ 
14 & ($-3.8,  4.4$) & 386.& 2.4 $\pm$ 0.5 & $ 2.1\times 1.6$ ($-59\arcdeg$) &  2.2 $\pm$  0.6 & 0.58 $\pm$ 0.04 &  1.3 $\pm$  0.7 \\ 
15 & ($-2.7,  4.6$) & 385.& 2.0 $\pm$ 2.0 & $ 2.5\times 1.3$ ($-4\arcdeg$) &\nodata& 0.10 $\pm$ 0.13 & \nodata \\ 
16 & ($ 8.7,  0.4$) & 384.& 2.2 $\pm$ 2.0 & $ 2.4\times 1.9$ ($59\arcdeg$) &  3.1 $\pm$  3.5 & 0.10 $\pm$ 0.18 &  1.5 $\pm$  3.0 \\ 
17 & ($-4.0,  1.6$) & 387.& 6.3 $\pm$ 5.3 & $ 7.0\times 1.7$ ($73\arcdeg$) &  4.9 $\pm$  4.4 & 0.47 $\pm$ 0.68 & 20.0 $\pm$ 42.1 \\ 
18 & ($-4.4,  4.5$) & 386.& 2.4 $\pm$ 2.9 & $ 2.9\times 1.7$ ($13\arcdeg$) &  3.1 $\pm$  2.7 & 0.20 $\pm$ 0.19 &  1.8 $\pm$  5.4 \\ 
19 & ($-0.8,  1.9$) & 388.& 2.2 $\pm$ 0.5 & $ 2.3\times 1.8$ ($-85\arcdeg$) &  2.8 $\pm$  0.9 & 0.37 $\pm$ 0.06 &  1.4 $\pm$  0.8 \\ 
20 & ($ 5.0, -6.4$) & 389.& 2.1 $\pm$ 1.5 & $ 6.6\times 1.5$ ($20\arcdeg$) &  3.7 $\pm$  2.2 & 0.22 $\pm$ 0.19 &  1.6 $\pm$  2.4 \\ 
21 & ($-3.5,  3.2$) & 389.& 2.2 $\pm$ 0.9 & $ 3.9\times 1.3$ ($-52\arcdeg$) &\nodata& 0.24 $\pm$ 0.12 & \nodata \\ 
22 & ($-9.6, 10.3$) & 389.& 1.0 $\pm$ 0.4 & $ 5.2\times 2.5$ ($-43\arcdeg$) &  6.2 $\pm$  1.6 & 0.65 $\pm$ 0.15 &  0.7 $\pm$  0.5 \\ 
23 & ($ 6.4, 10.7$) & 388.& 1.4 $\pm$ 0.4 & $ 3.3\times 1.7$ ($-9\arcdeg$) &  3.5 $\pm$  1.0 & 0.28 $\pm$ 0.09 &  0.7 $\pm$  0.5 \\ 
24 & ($ 6.1, 11.0$) & 388.& 1.3 $\pm$ 0.6 & $ 2.5\times 1.9$ ($34\arcdeg$) &  3.2 $\pm$  1.3 & 0.14 $\pm$ 0.06 &  0.5 $\pm$  0.5 \\ 
25 & ($-1.7, 12.4$) & 388.& 0.7 $\pm$ 0.5 & $ 2.2\times 1.5$ ($-88\arcdeg$) &  1.9 $\pm$  1.3 & 0.13 $\pm$ 0.07 &  0.1 $\pm$  0.2 \\ 
26 & ($-2.8,  1.2$) & 390.& 1.5 $\pm$ 0.6 & $ 2.4\times 1.8$ ($36\arcdeg$) &  3.0 $\pm$  2.3 & 0.18 $\pm$ 0.13 &  0.7 $\pm$  0.8 \\ 
27 & ($ 0.3,  3.7$) & 389.& 2.7 $\pm$ 2.3 & $ 2.2\times 1.9$ ($-21\arcdeg$) &  2.9 $\pm$  3.3 & 0.10 $\pm$ 0.20 &  2.2 $\pm$  4.9 \\ 
28 & ($-0.6,  2.4$) & 397.& 3.1 $\pm$ 0.4 & $ 5.1\times 2.3$ ($-48\arcdeg$) &  5.7 $\pm$  0.6 & 1.45 $\pm$ 0.11 &  5.7 $\pm$  1.8 \\ 
29 & ($-0.5,  3.1$) & 393.& 2.0 $\pm$ 0.5 & $ 2.2\times 1.6$ ($-36\arcdeg$) &  2.3 $\pm$  0.8 & 0.36 $\pm$ 0.05 &  0.9 $\pm$  0.6 \\ 
30 & ($-4.4,  5.0$) & 393.& 1.8 $\pm$ 0.9 & $ 4.1\times 1.8$ ($-72\arcdeg$) &  4.0 $\pm$  1.7 & 0.25 $\pm$ 0.14 &  1.3 $\pm$  1.6 \\ 
31 & ($ 1.2,  7.0$) & 396.& 2.7 $\pm$ 0.5 & $ 3.1\times 2.6$ ($-65\arcdeg$) &  4.8 $\pm$  1.0 & 0.81 $\pm$ 0.09 &  3.7 $\pm$  1.3 \\ 
32 & ($-0.2,  7.2$) & 397.& 3.3 $\pm$ 0.3 & $ 3.4\times 3.0$ ($43\arcdeg$) &  5.6 $\pm$  0.7 & 1.53 $\pm$ 0.08 &  6.4 $\pm$  1.7 \\ 
33 & ($ 6.0,  9.8$) & 393.& 1.6 $\pm$ 0.5 & $ 3.6\times 1.9$ ($-9\arcdeg$) &  3.9 $\pm$  1.2 & 0.28 $\pm$ 0.08 &  1.1 $\pm$  0.7 \\ 
34 & ($-4.9,  2.2$) & 394.& 0.7 $\pm$ 0.4 & $ 2.1\times 1.5$ ($-63\arcdeg$) &  2.0 $\pm$  0.9 & 0.17 $\pm$ 0.04 &  0.1 $\pm$  0.1 \\ 
35 & ($-1.5,  3.7$) & 396.& 3.1 $\pm$ 1.8 & $ 1.9\times 1.6$ ($58\arcdeg$) &  2.0 $\pm$  2.3 & 0.18 $\pm$ 0.14 &  2.1 $\pm$  3.3 \\ 
36 & ($-6.6,  5.0$) & 396.& 2.3 $\pm$ 0.5 & $ 7.0\times 1.9$ ($-53\arcdeg$) &  5.8 $\pm$  1.4 & 0.72 $\pm$ 0.11 &  3.3 $\pm$  1.6 \\ 
37 & ($-7.1,  4.7$) & 395.& 2.8 $\pm$ 0.8 & $ 3.3\times 2.2$ ($60\arcdeg$) &  4.4 $\pm$  1.2 & 0.50 $\pm$ 0.11 &  3.7 $\pm$  2.5 \\ 
38 & ($ 2.3,  6.1$) & 397.& 2.1 $\pm$ 0.4 & $ 4.2\times 4.4$ ($-16\arcdeg$) &  7.8 $\pm$  1.0 & 1.15 $\pm$ 0.13 &  3.6 $\pm$  1.4 \\ 
39 & ($ 6.5,  8.2$) & 393.& 1.0 $\pm$ 0.5 & $ 2.0\times 1.8$ ($51\arcdeg$) &  2.5 $\pm$  1.2 & 0.17 $\pm$ 0.07 &  0.2 $\pm$  0.3 \\ 
40 & ($ 6.4,  7.9$) & 394.& 1.6 $\pm$ 0.8 & $ 1.6\times 1.1$ ($10\arcdeg$) &\nodata& 0.10 $\pm$ 0.06 & \nodata \\ 
41 & ($ 0.7,  3.1$) & 394.& 1.0 $\pm$ 0.5 & $ 4.0\times 1.3$ ($88\arcdeg$) &\nodata& 0.22 $\pm$ 0.07 & \nodata \\ 
42 & ($ 1.4,  4.5$) & 395.& 3.6 $\pm$ 2.5 & $ 2.1\times 1.5$ ($-23\arcdeg$) &  1.8 $\pm$  2.6 & 0.15 $\pm$ 0.11 &  2.5 $\pm$  4.9 \\ 
43 & ($ 0.7,  9.0$) & 394.& 1.4 $\pm$ 0.9 & $ 2.2\times 1.3$ ($-73\arcdeg$) &\nodata& 0.08 $\pm$ 0.08 & \nodata \\ 
44 & ($-0.9,  4.3$) & 398.& 2.8 $\pm$ 1.4 & $ 1.9\times 1.6$ ($20\arcdeg$) &  2.1 $\pm$  1.5 & 0.25 $\pm$ 0.13 &  1.7 $\pm$  2.4 \\ 
45 & ($-4.3,  5.4$) & 396.& 3.0 $\pm$ 1.5 & $ 1.5\times 1.3$ ($-3\arcdeg$) &\nodata& 0.13 $\pm$ 0.07 & \nodata \\ 
46 & ($-3.4,  5.7$) & 403.& 6.1 $\pm$ 3.2 & $ 5.4\times 1.8$ ($-45\arcdeg$) &  4.6 $\pm$  1.9 & 0.80 $\pm$ 0.39 & 17.7 $\pm$ 23.8 \\ 
47 & ($19.9, -6.2$) & 399.& 1.0 $\pm$ 0.6 & $ 5.5\times 2.7$ ($-79\arcdeg$) &  6.7 $\pm$  2.3 & 0.54 $\pm$ 0.19 &  0.7 $\pm$  0.9 \\ 
48 & ($-5.3,  3.4$) & 400.& 2.3 $\pm$ 1.9 & $ 2.2\times 1.4$ ($8\arcdeg$) &  1.5 $\pm$  1.6 & 0.10 $\pm$ 0.12 &  0.8 $\pm$  1.5 \\ 
49 & ($ 1.9,  5.0$) & 400.& 1.9 $\pm$ 1.1 & $ 2.8\times 2.3$ ($-88\arcdeg$) &  4.0 $\pm$  1.9 & 0.46 $\pm$ 0.25 &  1.6 $\pm$  2.0 \\ 
50 & ($-2.2, -7.6$) & 403.& 2.1 $\pm$ 0.6 & $ 1.8\times 1.5$ ($-74\arcdeg$) &  1.5 $\pm$  1.2 & 0.18 $\pm$ 0.07 &  0.7 $\pm$  0.7 \\ 
51 & ($18.2, -5.5$) & 402.& 1.5 $\pm$ 0.6 & $ 3.1\times 1.4$ ($89\arcdeg$) &  2.1 $\pm$  0.6 & 0.28 $\pm$ 0.06 &  0.5 $\pm$  0.4 \\ 
52 & ($-2.2, -1.1$) & 401.& 1.8 $\pm$ 1.7 & $ 3.0\times 1.5$ ($-25\arcdeg$) &  2.5 $\pm$  1.9 & 0.13 $\pm$ 0.11 &  0.8 $\pm$  2.0 \\ 
53 & ($-2.2,  3.1$) & 408.& 4.3 $\pm$ 0.5 & $ 4.3\times 2.5$ ($32\arcdeg$) &  5.6 $\pm$  0.6 & 1.50 $\pm$ 0.13 & 10.5 $\pm$  2.8 \\ 
54 & ($ 2.3,  3.4$) & 400.& 1.9 $\pm$ 1.6 & $ 3.0\times 1.6$ ($69\arcdeg$) &  2.7 $\pm$  2.6 & 0.14 $\pm$ 0.14 &  1.0 $\pm$  2.0 \\ 
55 & ($-0.7,  3.7$) & 405.& 3.6 $\pm$ 1.1 & $ 2.9\times 2.0$ ($65\arcdeg$) &  3.8 $\pm$  1.5 & 0.51 $\pm$ 0.10 &  5.1 $\pm$  3.3 \\ 
56 & ($-1.7, -0.9$) & 402.& 1.1 $\pm$ 2.0 & $ 1.9\times 1.8$ ($49\arcdeg$) &  2.5 $\pm$  3.3 & 0.06 $\pm$ 0.11 &  0.3 $\pm$  1.2 \\ 
57 & ($ 2.5,  3.7$) & 404.& 1.4 $\pm$ 0.5 & $ 2.0\times 1.5$ ($43\arcdeg$) &  2.0 $\pm$  1.2 & 0.18 $\pm$ 0.08 &  0.4 $\pm$  0.3 \\ 
58 & ($-0.2,  4.2$) & 402.& 1.4 $\pm$ 1.3 & $ 3.8\times 2.2$ ($-58\arcdeg$) &  4.7 $\pm$  3.0 & 0.20 $\pm$ 0.26 &  1.0 $\pm$  1.8 \\ 
59 & ($-8.9,  6.6$) & 402.& 3.1 $\pm$ 2.6 & $ 4.1\times 1.3$ ($81\arcdeg$) &\nodata& 0.16 $\pm$ 0.27 & \nodata \\ 
60 & ($18.6, -6.4$) & 404.& 3.4 $\pm$ 2.4 & $ 3.1\times 1.1$ ($-34\arcdeg$) &\nodata& 0.21 $\pm$ 0.15 & \nodata \\ 
61 & ($ 3.0,  4.2$) & 404.& 2.3 $\pm$ 1.9 & $ 3.2\times 1.3$ ($85\arcdeg$) &\nodata& 0.21 $\pm$ 0.23 & \nodata \\ 
62 & ($-3.5, -5.1$) & 405.& 2.6 $\pm$ 2.9 & $ 3.7\times 1.9$ ($19\arcdeg$) &  4.2 $\pm$  5.2 & 0.15 $\pm$ 0.27 &  3.0 $\pm$  9.2 \\ 
63 & ($-0.5, -4.7$) & 404.& 1.1 $\pm$ 0.9 & $ 2.7\times 1.3$ ($-61\arcdeg$) &\nodata& 0.09 $\pm$ 0.06 & \nodata \\ 
64 & ($-2.4,  4.4$) & 405.& 1.3 $\pm$ 0.9 & $ 1.9\times 1.5$ ($35\arcdeg$) &  1.6 $\pm$  1.7 & 0.10 $\pm$ 0.08 &  0.3 $\pm$  0.4 \\ 
65 & ($ 0.2,  6.4$) & 405.& 2.8 $\pm$ 1.8 & $ 2.7\times 1.2$ ($-38\arcdeg$) &\nodata& 0.11 $\pm$ 0.12 & \nodata \\ 
66 & ($-11.3,  7.7$) & 407.& 2.0 $\pm$ 0.4 & $ 2.8\times 2.4$ ($-61\arcdeg$) &  4.2 $\pm$  0.8 & 0.43 $\pm$ 0.08 &  1.7 $\pm$  0.8 \\ 
67 & ($-1.5, -4.2$) & 407.& 1.0 $\pm$ 0.4 & $ 2.3\times 1.5$ ($17\arcdeg$) &  2.0 $\pm$  1.0 & 0.15 $\pm$ 0.07 &  0.2 $\pm$  0.2 \\ 
68 & ($15.6, -3.7$) & 407.& 1.4 $\pm$ 0.9 & $ 4.4\times 2.6$ ($-27\arcdeg$) &  5.8 $\pm$  2.3 & 0.35 $\pm$ 0.16 &  1.1 $\pm$  1.6 \\ 
69 & ($-7.6,  2.4$) & 407.& 1.7 $\pm$ 1.8 & $ 2.7\times 1.8$ ($-89\arcdeg$) &  3.2 $\pm$  2.3 & 0.08 $\pm$ 0.11 &  0.9 $\pm$  2.2 \\ 
70 & ($11.1, -6.3$) & 412.& 2.1 $\pm$ 0.9 & $ 3.9\times 3.0$ ($-14\arcdeg$) &  5.9 $\pm$  2.8 & 0.81 $\pm$ 0.37 &  2.7 $\pm$  2.8 \\ 
71 & ($ 5.3, -6.6$) & 411.& 1.2 $\pm$ 0.7 & $ 2.3\times 1.6$ ($-10\arcdeg$) &  2.4 $\pm$  1.5 & 0.11 $\pm$ 0.07 &  0.3 $\pm$  0.4 \\ 
72 & ($10.0, -3.9$) & 420.& 7.0 $\pm$ 0.5 & $15.4\times 7.9$ ($61\arcdeg$) & 20.9 $\pm$  1.7 & 5.50 $\pm$ 0.38 & 106.9 $\pm$ 21.2 \\ 
73 & ($ 4.8, -8.7$) & 413.& 1.4 $\pm$ 0.6 & $ 3.4\times 2.0$ ($-10\arcdeg$) &  4.1 $\pm$  1.6 & 0.23 $\pm$ 0.06 &  0.8 $\pm$  0.8 \\ 
74 & ($ 5.3, -8.8$) & 414.& 2.8 $\pm$ 1.8 & $ 3.3\times 2.4$ ($-47\arcdeg$) &  4.7 $\pm$  2.2 & 0.21 $\pm$ 0.14 &  3.8 $\pm$  6.2 \\ 
75 & ($-0.7,  3.3$) & 413.& 1.9 $\pm$ 0.8 & $ 2.0\times 1.4$ ($60\arcdeg$) &  0.9 $\pm$  0.8 & 0.17 $\pm$ 0.06 &  0.3 $\pm$  0.4 \\ 
76 & ($10.2, -5.7$) & 414.& 3.3 $\pm$ 1.7 & $ 3.9\times 1.6$ ($-73\arcdeg$) &  3.4 $\pm$  2.3 & 0.25 $\pm$ 0.25 &  3.9 $\pm$  5.5 \\ 
77 & ($ 9.9, -5.3$) & 416.& 3.6 $\pm$ 1.6 & $ 4.5\times 3.4$ ($-66\arcdeg$) &  7.1 $\pm$  2.3 & 0.52 $\pm$ 0.21 &  9.3 $\pm$  9.6 \\ 
78 & ($12.4, -2.5$) & 413.& 1.6 $\pm$ 1.8 & $ 2.2\times 1.2$ ($-81\arcdeg$) &\nodata& 0.09 $\pm$ 0.11 & \nodata \\ 
79 & ($10.0, -1.8$) & 415.& 1.8 $\pm$ 1.3 & $ 6.1\times 3.7$ ($-26\arcdeg$) &  8.6 $\pm$  4.6 & 1.23 $\pm$ 0.55 &  3.0 $\pm$  4.5 \\ 
80 & ($ 9.0, -5.0$) & 416.& 1.5 $\pm$ 0.9 & $ 2.7\times 1.5$ ($7\arcdeg$) &  2.4 $\pm$  1.5 & 0.09 $\pm$ 0.10 &  0.5 $\pm$  0.8 \\ 
81 & ($ 7.1, -2.0$) & 419.& 2.5 $\pm$ 0.4 & $ 8.7\times 5.0$ ($6\arcdeg$) & 12.2 $\pm$  2.4 & 3.50 $\pm$ 0.49 &  7.8 $\pm$  2.6 \\ 
82 & ($ 7.8, -1.9$) & 416.& 1.8 $\pm$ 2.6 & $ 3.1\times 3.6$ ($13\arcdeg$) &  5.8 $\pm$  9.9 & 0.53 $\pm$ 0.58 &  1.9 $\pm$  7.5 \\ 
83 & ($12.2, -5.9$) & 422.& 2.6 $\pm$ 0.8 & $ 3.6\times 2.2$ ($-77\arcdeg$) &  4.7 $\pm$  1.1 & 0.51 $\pm$ 0.16 &  3.3 $\pm$  2.5 \\ 
84 & ($ 4.3, -2.6$) & 423.& 2.2 $\pm$ 0.3 & $ 6.4\times 5.1$ ($-13\arcdeg$) & 10.6 $\pm$  1.1 & 1.65 $\pm$ 0.12 &  5.2 $\pm$  1.9 \\ 
85 & ($ 3.5, -0.8$) & 421.& 1.1 $\pm$ 1.8 & $ 4.8\times 2.2$ ($45\arcdeg$) &  5.3 $\pm$  8.1 & 0.13 $\pm$ 0.22 &  0.7 $\pm$  2.9 \\ 
86 & ($ 0.0, -7.9$) & 422.& 0.5 $\pm$ 0.1 & $ 2.2\times 1.9$ ($15\arcdeg$) &  3.0 $\pm$  1.5 & 0.19 $\pm$ 0.07 &  0.1 $\pm$  0.1 \\ 
87 & ($-0.9, -7.1$) & 422.& 1.1 $\pm$ 0.8 & $ 1.8\times 1.6$ ($34\arcdeg$) &  1.9 $\pm$  1.9 & 0.09 $\pm$ 0.07 &  0.2 $\pm$  0.4 \\ 
88 & ($10.9, -6.1$) & 423.& 2.0 $\pm$ 1.6 & $ 3.4\times 2.8$ ($-43\arcdeg$) &  5.2 $\pm$  3.1 & 0.21 $\pm$ 0.25 &  2.1 $\pm$  3.9 \\ 
89 & ($10.7, -5.8$) & 422.& 0.7 $\pm$ 1.4 & $ 2.5\times 1.5$ ($-88\arcdeg$) &  2.2 $\pm$  2.5 & 0.07 $\pm$ 0.15 &  0.1 $\pm$  0.5 \\ 
90 & ($ 1.9, -1.6$) & 426.& 3.0 $\pm$ 0.5 & $ 5.0\times 4.1$ ($-44\arcdeg$) &  8.3 $\pm$  0.8 & 1.41 $\pm$ 0.22 &  7.6 $\pm$  2.8 \\ 
91 & ($ 8.4, -5.2$) & 424.& 1.4 $\pm$ 1.0 & $ 3.3\times 1.7$ ($-5\arcdeg$) &  3.4 $\pm$  2.0 & 0.19 $\pm$ 0.11 &  0.7 $\pm$  1.2 \\ 
92 & ($ 5.9, -3.6$) & 426.& 1.6 $\pm$ 0.2 & $ 7.7\times 5.2$ ($-51\arcdeg$) & 11.8 $\pm$  1.0 & 1.85 $\pm$ 0.18 &  3.2 $\pm$  0.9 \\ 
93 & ($10.5, -10.6$) & 428.& 2.4 $\pm$ 2.1 & $ 2.2\times 1.7$ ($-72\arcdeg$) &  2.5 $\pm$  2.4 & 0.20 $\pm$ 0.30 &  1.6 $\pm$  3.3 \\ 
94 & ($ 0.1, -8.5$) & 426.& 2.3 $\pm$ 2.1 & $ 4.0\times 1.6$ ($5\arcdeg$) &  3.5 $\pm$  4.1 & 0.16 $\pm$ 0.26 &  1.9 $\pm$  4.6 \\ 
95 & ($14.9, -6.9$) & 425.& 0.7 $\pm$ 1.1 & $ 4.4\times 1.9$ ($-60\arcdeg$) &  4.4 $\pm$  3.4 & 0.16 $\pm$ 0.36 &  0.2 $\pm$  0.7 \\ 
96 & ($14.5, -6.6$) & 426.& 1.4 $\pm$ 2.3 & $ 4.9\times 3.0$ ($-1\arcdeg$) &  6.8 $\pm$  5.0 & 0.28 $\pm$ 0.28 &  1.3 $\pm$  4.6 \\ 
97 & ($10.0, -5.8$) & 427.& 1.8 $\pm$ 1.0 & $ 2.2\times 2.1$ ($77\arcdeg$) &  3.3 $\pm$  1.9 & 0.25 $\pm$ 0.12 &  1.1 $\pm$  1.3 \\ 
98 & ($ 8.9, -5.7$) & 425.& 1.2 $\pm$ 0.7 & $ 2.2\times 1.4$ ($-54\arcdeg$) &  1.3 $\pm$  1.1 & 0.11 $\pm$ 0.07 &  0.2 $\pm$  0.3 \\ 
99 & ($11.2, -5.4$) & 427.& 1.9 $\pm$ 1.5 & $ 3.1\times 1.9$ ($87\arcdeg$) &  3.6 $\pm$  3.1 & 0.22 $\pm$ 0.20 &  1.3 $\pm$  2.6 \\ 
100 & ($ 5.0, -4.6$) & 427.& 1.6 $\pm$ 0.7 & $ 2.4\times 1.7$ ($2\arcdeg$) &  2.7 $\pm$  1.0 & 0.27 $\pm$ 0.08 &  0.7 $\pm$  0.7 \\ 
101 & ($ 8.0, -5.4$) & 428.& 0.9 $\pm$ 0.4 & $ 5.5\times 1.5$ ($-72\arcdeg$) &  3.7 $\pm$  1.4 & 0.32 $\pm$ 0.09 &  0.3 $\pm$  0.3 \\ 
102 & ($16.2, -5.4$) & 428.& 3.7 $\pm$ 3.5 & $ 3.7\times 2.4$ ($-23\arcdeg$) &  5.0 $\pm$  4.4 & 0.19 $\pm$ 0.41 &  7.1 $\pm$ 13.7 \\ 
103 & ($10.9, -4.1$) & 426.& 2.8 $\pm$ 4.0 & $ 3.2\times 2.7$ ($-29\arcdeg$) &  4.9 $\pm$  6.4 & 0.13 $\pm$ 0.36 &  4.0 $\pm$ 12.0 \\ 
104 & ($ 7.8, -3.8$) & 428.& 1.3 $\pm$ 0.7 & $ 1.7\times 1.4$ ($28\arcdeg$) &  1.0 $\pm$  1.6 & 0.09 $\pm$ 0.06 &  0.2 $\pm$  0.2 \\ 
105 & ($ 2.7, -2.2$) & 428.& 0.5 $\pm$ 0.4 & $ 2.4\times 2.2$ ($41\arcdeg$) &  3.6 $\pm$  2.5 & 0.13 $\pm$ 0.16 &  0.1 $\pm$  0.2 \\ 
106 & ($ 4.0, -6.9$) & 432.& 1.3 $\pm$ 1.1 & $ 2.8\times 1.8$ ($42\arcdeg$) &  3.3 $\pm$  1.7 & 0.15 $\pm$ 0.13 &  0.5 $\pm$  1.0 \\ 
107 & ($ 4.3, -5.5$) & 433.& 2.6 $\pm$ 1.0 & $ 3.2\times 1.7$ ($64\arcdeg$) &  3.4 $\pm$  1.5 & 0.25 $\pm$ 0.12 &  2.5 $\pm$  2.5 \\ 
108 & ($ 6.3, -5.1$) & 433.& 0.7 $\pm$ 0.4 & $ 3.6\times 1.7$ ($47\arcdeg$) &  3.5 $\pm$  1.1 & 0.20 $\pm$ 0.09 &  0.2 $\pm$  0.2 \\ 
109 & ($ 7.0, -4.9$) & 434.& 0.5 $\pm$ 0.5 & $ 4.6\times 1.7$ ($57\arcdeg$) &  4.0 $\pm$  2.2 & 0.17 $\pm$ 0.17 &  0.1 $\pm$  0.2 \\ 
110 & ($ 6.8, -2.9$) & 432.& 1.8 $\pm$ 0.6 & $ 4.3\times 2.2$ ($54\arcdeg$) &  5.1 $\pm$  2.1 & 0.28 $\pm$ 0.16 &  1.8 $\pm$  1.5 \\ 
111 & ($ 3.3, -0.2$) & 436.& 3.8 $\pm$ 0.7 & $ 2.9\times 2.0$ ($-31\arcdeg$) &  3.8 $\pm$  0.8 & 0.75 $\pm$ 0.10 &  5.6 $\pm$  2.7 \\ 
112 & ($-0.6, -8.6$) & 441.& 1.1 $\pm$ 0.5 & $ 3.0\times 1.5$ ($44\arcdeg$) &  2.4 $\pm$  0.8 & 0.22 $\pm$ 0.08 &  0.3 $\pm$  0.3 \\ 
113 & ($-1.0, -8.5$) & 442.& 1.5 $\pm$ 0.8 & $ 2.3\times 1.5$ ($-62\arcdeg$) &  2.2 $\pm$  1.4 & 0.09 $\pm$ 0.07 &  0.5 $\pm$  0.6 \\ 
114 & ($-1.1, -7.8$) & 440.& 1.8 $\pm$ 4.8 & $ 3.5\times 1.9$ ($38\arcdeg$) &  3.9 $\pm$  6.3 & 0.12 $\pm$ 0.31 &  1.3 $\pm$  6.9 \\ 
115 & ($-1.1, -7.2$) & 439.& 1.8 $\pm$ 1.2 & $ 2.0\times 1.5$ ($-41\arcdeg$) &  1.7 $\pm$  1.7 & 0.11 $\pm$ 0.08 &  0.6 $\pm$  1.0 \\ 
116 & ($-1.7, -5.6$) & 445.& 2.4 $\pm$ 0.5 & $ 2.6\times 1.3$ ($36\arcdeg$) &\nodata& 0.31 $\pm$ 0.07 & \nodata \\ 
117 & ($ 0.3, -9.6$) & 445.& 1.2 $\pm$ 0.4 & $ 1.9\times 1.8$ ($-7\arcdeg$) &  2.4 $\pm$  1.0 & 0.20 $\pm$ 0.04 &  0.3 $\pm$  0.3 \\ 
118 & ($ 3.0, -2.2$) & 448.& 0.9 $\pm$ 0.7 & $ 2.2\times 1.4$ ($-84\arcdeg$) &  1.3 $\pm$  1.0 & 0.09 $\pm$ 0.07 &  0.1 $\pm$  0.2 
\enddata
\tablenotetext{a}{Intensity-weighted peak position relative to the center at the optically defined galaxy center, $\alpha=13^{\rm h}39^{\rm m}56\fs041$, $\delta=-31\arcdeg38\arcmin30\farcs03$.}
\tablenotetext{b}{Major and minor axes of the clouds without beam deconvolution. The position angles are indicated in parentheses (measured counterclockwise from north to east).}
\tablenotetext{c}{Radius and virial masses are not shown for the cloud whose minor axis is too small to calculate a deconvolved minor axis.}
\end{deluxetable*}

\clearpage

\begin{deluxetable}{ccccccc}
\tablecaption{Clouds and Their associated star clusters \label{sfe}}
\tablewidth{0pt}
\tablehead{
\colhead{Cloud} & \colhead{Cluster} & \colhead{Age} & \colhead{$M_{\ast}$}& \colhead{SFE} & \colhead{$F_{\rm rad}$} & \colhead{$F_{\rm grav}$\tablenotemark{(d)}}  \\
\colhead{} & \colhead{} & \colhead{(Myr)} & \colhead{($10^4\,\Msol$)}& \colhead{}& \colhead{($10^{30}$\,dynes)}& \colhead{($10^{30}$\,dynes)}
}
\startdata
5&  G-125, G-127, H-2$^{\ast}$ (G-129)           &  10\tablenotemark{(c)} & 4.6   & 0.64  & $<7.7$ & 1.1 (2.9) \\\hline
8 & C-5$^{\ast}$ (G-87,H-1),C-11$^{\ast}$   & $1^{+1}_{-1}$, $1^{+1}_{-1}$ & $7.46^{+0.20}_{-0.27}$,$25.5^{+6.7}_{-4.2}$& $0.40$  & 55.0 & 52.1 (84.1)  \\  \hline  %
28& C-3$^{\ast}$ (G-106, H-8) &$5^{+1}_{-2}$ & $0.46^{+0.11}_{-0.10}$ & 0.05 & 0.8 & 6.1 (6.4)  \\\hline
\multirow{2}{*}{31}& \multirow{2}{*}{G-105$^{\ast}$} & $2.0^{+0.1}_{-2.0}$\tablenotemark{(a)} & $0.79^{+0.07}_{-0.06}$\tablenotemark{(a)} & 0.15  & 1.3 & 2.7 (3.2)  \\     
    &                          & $4.5^{+31.0}_{-0.5}$\tablenotemark{(b)}  & $0.98^{+0.50}_{-0.59}$\tablenotemark{(b)} & 0.18 & 1.6 & 2.7 (3.3)  \\ \hline  %GALEV
53& C-4$^{\ast}$, G-86  & $6^{+0}_{-2}$ & $1.62^{+0.52}_{-0.48}$   &  0.16  & $<2.7$ &6.8 (8.1)\\\hline
\multirow{2}{*}{59}& \multirow{2}{*}{G-9$^{\ast}$ }    & $7.9^{+71.5}_{-1.9}$\tablenotemark{(a)}  & $0.13^{+0.14}_{-0.05}$\tablenotemark{(a)} & 0.13 & $<0.2$ & $<0.1$ ($<0.1$)  \\  
    &      & $100^{+12}_{-11}$\tablenotemark{(b)}  & $0.93^{+0.57}_{-0.03}$\tablenotemark{(b)} & 0.51  &$<1.6$ &$<0.1$  ($<0.2$)     %GALEV
\enddata
\tablecomments{ Column 1: cloud ID. Column 2: star cluster ID. The asterisk ($\ast$) symbol indicate a star cluster   C-$n$ from \citet{2015ApJ...811...75C}, G-$n$ from \citet{2013MNRAS.431.2917D}, H-$n$ from \citet{2004ApJ...603..503H}. Column 3: estimated age of the star cluster with the asterisk symbol in Column 2. If it is the estimated age in \citet{2013MNRAS.431.2917D}, two estimated ages and masses derived from two different models are listed. Column 4: estimated mass of the star cluster with the asterisk symbol in Column 2.  Column 5: star formation efficiency. Column 6: force due to radiation. Column 7: gravitational force. For the definition, see the text.}
\tablenotetext{(a)}{The age and mass are estimated using the {\sf YGGDRASIL} model \citep[see ][for details]{2013MNRAS.431.2917D}. }
\tablenotetext{(b)}{The age and mass are estimated using the {\sf GALEV} model \citep[see ][for details]{2013MNRAS.431.2917D}. }
\tablenotetext{(c)}{The age is estimated from the H$\alpha$ equivalent width in \citet{2004ApJ...603..503H}.}
\tablenotetext{(d)}{The $F_{\rm grav}$ calculated using ($M_{\rm gas}+M_{\ast}$) instead of $M_{\rm gas}$ is shown in parentheses.}
\end{deluxetable}

\begin{deluxetable}{lcccc}
\tablecaption{Properties of the structures showing smooth velocity gradients outside galaxy rotation\label{outflow}}
\tablewidth{0pt}
\tablehead{
\colhead{PA} & \colhead{Position\tablenotemark{a}} & \colhead{Mass} & \colhead{Velocity gradient}& \colhead{Gas flow rate\tablenotemark{b}}  \\
\colhead{} & \colhead{(arcsec)} & \colhead{($\times10^5$\Msol)} & \colhead{(\kms\,arcsec$^{-1}$)}& \colhead{(\Msol\,yr$^{-1}$)}
}
\startdata
70\arcdeg& $+4.6$ &4.6 & $-2.29$ & 0.07\\
100\arcdeg& $+5.6$ &3.3 & $+2.87$ &0.06\\
175\arcdeg& $-4.5$ &6.9 & $+1.34$ & 0.06
\enddata
\tablenotetext{a}{Offset from the center.}
\tablenotetext{b}{We assume an inclination of 45\arcdeg\ with respect to the line of sight for all structures.}
\end{deluxetable}

%\bibliographystyle{apj}
%\bibliography{bibdata}

\end{document}